\documentclass[twocolumn]{aastex63}
\usepackage{amsmath}

\newcommand{\ST}{\textsc{StarTrack}}
\newcommand{\msun}{M_\sun}
\newcommand{\rsun}{R_\sun}

\newcommand{\kmsMpc}{\mathrm{km\,s^{-1} Mpc^{-1}}}
\newcommand{\fbin}{f_\mathrm{bin}}
\usepackage{multirow}
\defcitealias{karczmarek22}{Paper~I}

\received{December 17, 2022}
\revised{March 22, 2023}
\accepted{March 26, 2023}
\shorttitle{The effect of binary Cepheids on the extragalactic distance scale}
\shortauthors{Karczmarek et al.}

\begin{document}

\title{Synthetic Population of Binary Cepheids. II. The effect of companion light on the extragalactic distance scale}

\author[0000-0002-0136-0046]{Paulina Karczmarek}
\affiliation{Departamento de Astronom{\'i}a, Universidad de Concepci{\'o}n, Casilla 160-C, Concepci{\'o}n, Chile}
\affiliation{Nicolaus Copernicus Astronomical Center, Bartycka 18, 00-716 Warsaw, Poland}

\author{Gergely Hajdu}
\affiliation{Nicolaus Copernicus Astronomical Center, Bartycka 18, 00-716 Warsaw, Poland}

\author{Grzegorz Pietrzy{\'n}ski}
\affiliation{Nicolaus Copernicus Astronomical Center, Bartycka 18, 00-716 Warsaw, Poland}

\author{Wolfgang Gieren}
\affiliation{Departamento de Astronom{\'i}a, Universidad de Concepci{\'o}n, Casilla 160-C, Concepci{\'o}n, Chile}

\author{Weronika Narloch}
\affiliation{Departamento de Astronom{\'i}a, Universidad de Concepci{\'o}n, Casilla 160-C, Concepci{\'o}n, Chile}

\author{Rados{\l}aw Smolec}
\affiliation{Nicolaus Copernicus Astronomical Center, Bartycka 18, 00-716 Warsaw, Poland}

\author{Grzegorz Wiktorowicz}
\affiliation{Nicolaus Copernicus Astronomical Center, Bartycka 18, 00-716 Warsaw, Poland}

\author{Krzysztof Belczynski}
\affiliation{Nicolaus Copernicus Astronomical Center, Bartycka 18, 00-716 Warsaw, Poland}


\correspondingauthor{Paulina Karczmarek}
\email{pkarczmarek@astro-udec.cl}



\begin{abstract}

Because of their period-luminosity relation (PLR), classical Cepheids play a key role in the calibration of the extragalactic distance scale and the determination of the Hubble-Lema\^{i}tre constant $H_0$.
Recent findings show that the majority of classical Cepheids should be in binary or multiple systems, which might undermine their accuracy, as the extra---and unaccounted for---light from the companions of Cepheids causes a shift in the PLR.
We quantify this shift using synthetic populations of binary Cepheids that we developed for this purpose, as described in Paper I of this series.
We find that while all PLRs are shifted toward brighter values due to the excess light from the companions, the bias in the relative distance modulus between two galaxies hosting binary Cepheids can be either positive or negative, depending on the percentage of binary Cepheids in them. If the binarity percentage in the two galaxies is similar, the effect of binarity is canceled. Otherwise, it introduces a shift in the distance modulus of the order of millimags in the near-infrared passbands and Wesenheit indices, and tens of millimags in the visual domain; its exact value depends on the variant of the synthetic population (a unique combination of metallicity, star formation history, shape and location of the instability strip, and initial parameter distributions).
Such shifts in distance moduli to type Ia supernova host galaxies introduce an additional statistical error on $H_0$, which however does not prevent measuring $H_0$ with a precision of 1\%.
\end{abstract}




\section{Introduction}
\label{sec:intro}

Classical Cepheids (hereafter referred to as Cepheids) are a vital component of the extragalactic distance scale. These young, massive (above $2.5 \msun$), and luminous pulsators follow tightly the period-luminosity relation (PLR), which makes them prominent and potent standard candles up to distances of $\sim 40$ megaparsecs \citep{riess16}.

Still, the accuracy of the PLR can be improved with a better understanding of factors that affect the intrinsic or apparent brightness of Cepheids, e.g. the effect of metallicity or binarity. Unlike the effect of metallicity \citep{breuval22}, the effect of binarity has so far been only described qualitatively, estimated based on a simplistic model, or assumed negligible \citep[e.g.][]{lanoix99,szabados12,Anderson16ApJS}.

Diligent quantification of the effect of binarity is indeed a challenging task, because the census of binary Cepheids is far from complete. Out of more than 3000 Milky Way (MW) Cepheids known so far \citep{pietrukowicz21} only about 200 have confirmed companions \citep[][Shetye et al. in prep.]{szabados03db,kervella19I}, which constitutes $\sim6$\% of all known MW Cepheids, yet $60-80$\% of them is expected to be binaries \citep{szabados03,neilson15,mor17}. In the Large and Small Magellanic Clouds (LMC, SMC), hosting altogether around 10000 Cepheids \citep{soszynski15ccep}, a total of about 80 binary Cepheids are known \citep{szabados12,pilecki21}. Such scarcity of observed binary Cepheids relative to the expected ones indicates a strong selection bias, which originates from the limited applications of the various binary detection techniques.

The evidence is growing that Cepheids are not only members of binary systems, but also triples and possibly quadruples \citep{evans05,evans20,dinnbier22}. This is causing a paradigm shift in our thinking about Cepheids first as single, later as binary, and now as triple and quadruple systems, and calls for action to include the presence of companions in the calculations of Cepheids' masses or apparent brightnesses, as neglecting the effect of companion(s) can lead to inaccurate estimations and constraints on stellar models.

Binary (and multiple) Cepheids appear brighter than their single counterparts, due to the extra---and unaccounted for---light of their companions. In the context of distance determination, Cepheids' companions contaminate the PLR, causing the zero point to ``shift upwards'' (toward brighter, i.e. more negative magnitudes). Consequently, the distance determined from the PLR of binary Cepheid seems smaller than if determined from the PLR of only single Cepheids. Binary Cepheids could have an adverse effect also on the slope of the PLR \citep{szabados12}, further distorting the calibration of the extragalactic distance scale. 

Since the majority of binary Cepheids evade detection, let alone elimination from the calibration and application of the PLR, we have explored an alternative, simulation-based approach, in order to estimate their effect on the extragalactic distance scale. We employed the binary population synthesis method, and created a collection of synthetic populations of binary Cepheids, whose properties we described and compared with observational data \citep[][hereafter \citetalias{karczmarek22}]{karczmarek22}. Synthetic data are by definition free from selection and completeness biases, unaffected by the reddening, and their crucial features, like metallicity and binarity, are established a priori. Therefore, synthetic data are advantageous in situations in which the real data are incomplete or inaccurate. In \citetalias{karczmarek22} of this series we established the synthetic population of binary Cepheids as a reliable tool to investigate the effect of the companions on the PLR, and in this work we provide a quantitative description of this effect on the extragalactic distance scale.

In section \ref{sec:syntpop} we recap the most important features of the synthetic populations that were extensively analyzed in \citetalias{karczmarek22}. In section \ref{sec:DeltaSLZP_bin} we present the effect of binarity on the slope and zero point of Cepheid PLRs, within a chosen metallicity environment. In section \ref{sec:DeltaDM_bin} we quantify the effect of binarity between the relative distance moduli of galaxies of different metallicities. We discuss the implications of binary Cepheids on $H_0$ in section \ref{sec:H0}, and in section \ref{sec:discussion} we address other, more subtle factors that might affect the results from sections \ref{sec:DeltaSLZP_bin} and \ref{sec:DeltaDM_bin}.

\section{Synthetic populations extended}
\label{sec:syntpop}

In \citetalias{karczmarek22} we created a collection of 48 synthetic populations of binary Cepheids, by applying the \ST\ population synthesis code \citep{belczynski02,belczynski08} to four different sets of initial parameter distributions (see Table 1 and Figure A1 in \citetalias{karczmarek22}), two prescriptions for the shape and location of the instability strip (IS), two star formation histories (SFH), and three metallicities $Z = \{ 0.004, 0.008, 0.02\}$, which reflect the mean metal content of Cepheids in the SMC, LMC, and MW, respectively (see Section 2 in \citetalias{karczmarek22}). The produced variants differed in a number of features. For instance, the relative fraction of Cepheids on their first instability strip crossing (IS1) is the largest for variants with uniform SFH, the number of hot Cepheid companion (spectral type BA) main sequence (MS) stars is the largest in variants with the SFH based on the observed age distribution of the Cepheids; for a detailed comparison between all the variants see \citetalias{karczmarek22}.

By analyzing different variants of binary Cepheid populations, we can assess the relevance of input parameters on the results. The differences between the results, arising from different variants, can also be adopted as a kind of systematic error and included in the total error budget. Moreover, a variant that yields results which are in obvious disagreement with observations, can be discarded as unrealistic, further constraining the input values and the results. For example, we have established that variants with initial parameter distributions from set B yield nonphysically high estimation on the percentage of binary Cepheids in the LMC (Section 4.4 in \citetalias{karczmarek22}), and variants with a uniform SFH show implausibly high percentage of IS1 Cepheids (Section 3.2 in \citetalias{karczmarek22}). The populations created from set B of the initial parameter distributions and the uniform SFH very likely do not reflect the true populations of binary Cepheids, nevertheless, we keep them in our analysis for the sake of completeness.

\begin{figure*}
    \centering
    \includegraphics[width=0.92\textwidth]{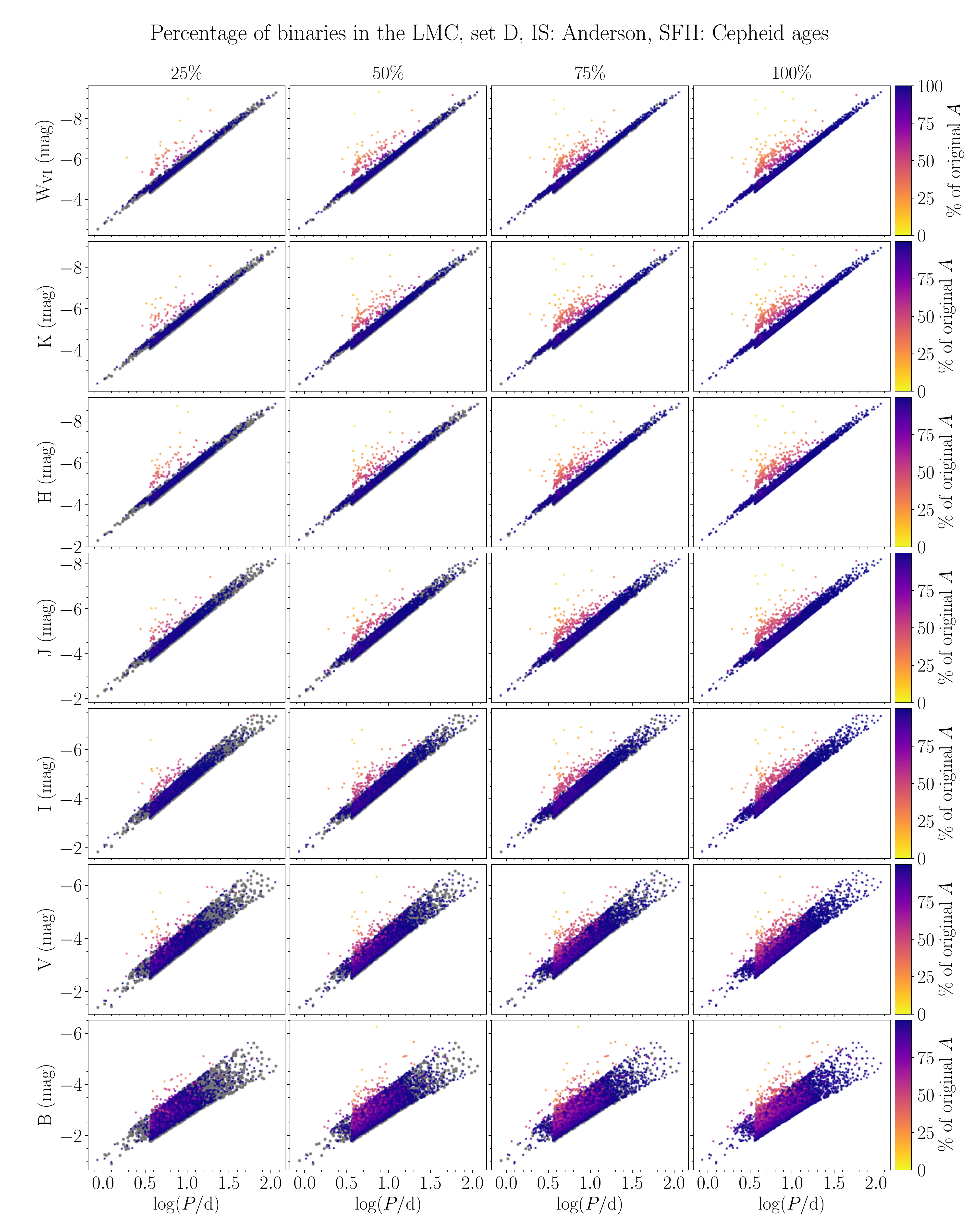}
    \caption{Multi-band $BVIJHKW_{VI}$ (bottom to top) period-luminosity relations for synthetic Cepheids with binarity fractions 25, 50, 75, and 100\% (left to right). Color-coded (dark to bright) are the relative pulsation amplitudes of Cepheids, diminished by the light of the companions with respect to the original pulsation amplitudes of single Cepheids, i.e. $100\% \times A_\mathrm{tot}/A_\mathrm{Cep}$. Grey points mark single Cepheids.}
    \label{fig:percentLMC}
\end{figure*}

As already introduced in \citetalias{karczmarek22}, the effect of binarity is controlled by the \emph{binarity percentage}, $\fbin$, which is a free parameter that can range from 0\% (only single Cepheids in the population), through 25, 50, 75, to 100\% (i.e., all Cepheids are in binaries). Now we explain further how a synthetic population with a given $\fbin$ is created and illustrate the results in Figure \ref{fig:percentLMC}.

By default, every synthetic population in our collection consists of 10000 binary systems ($\fbin=100\%$.) Every system has passed the filtering algorithm (described in detail in \citetalias{karczmarek22}, Section 2), which selected only stars that fulfilled the requirements for Cepheid variables and had not experienced substantial mass transfer or a common envelope episode. The fact that Cepheids evolved without interactions with their companions allows us to ignore the existence of companions for the sake of creating a mixed population of single+binary Cepheids. In order to do that, we first select the binarity percentage, for instance $\fbin=50\%$. Next, choose 50\% of the systems randomly in our population for which we simply ignore the extra light from the companion. As a result, the total brightness of these systems comes solely from the Cepheids. For the remaining 50\% of the population, for which the extra light from the companion is taken into account, we calculate the total magnitude of each system as:
\begin{equation}
    m_{\rm tot} = -2.5 \log \left( 10^{-m_A/2.5}  + 10^{-m_B/2.5}\right)\,,
\end{equation}
where $m_A$ and $m_B$ are the magnitudes of the primary and secondary, respectively.

Figure \ref{fig:percentLMC} illustrates the PLRs of the LMC synthetic Cepheids with companions that constitute 25, 50, 75, and 100\% of the sample. Plots present one variant of the synthetic population as an example, composed of binaries created from set D of the initial parameter distributions, filtered with the IS prescription of \citet{anderson16} and scaled to the SFH based on the age distribution of the Cepheids. This variant was chosen because it seems to best emulate the true population of Cepheids in terms of the IS and SFH, while set D incorporates the most up-to-date research on the initial parameter distributions of binary stars.

Grey data points in Figure \ref{fig:percentLMC} represent single Cepheids, with the rest being binary Cepheids. The resulting fractional pulsation amplitude (diminished due to the light from the Cepheid companion) is encoded in the color of the points, with the darkest points being almost unaffected Cepheids due to the minimal contribution of the companions to the total light of the system. Meanwhile, the amplitude of the Cepheids denoted by the brightest points is diminished to just a few percent of their original values. A gap in the data at $\log (P/\mathrm{d}) \approx 0.6$ separates IS1 from IS2+3 Cepheids. IS1 Cepheids are presented for the sake of completeness, since they met all the requirements for pulsating stars set in the filtering algorithm, but their counts, and properties demand further investigation, which is beyond the scope of this paper.

The scatter of the PLRs in Figure \ref{fig:percentLMC} is due to the finite width of the IS (the largest for the shortest wavelengths, the smallest for the longest wavelengths, and the Wesenheit index $W_{VI}$), but also due to the binary nature of Cepheids. The biggest contributors to that scatter are bright, giant companions located well above the main period-luminosity trend, and best visible in the near-infrared (NIR) and $W_{VI}$. They are easily detectable because they are 1.5-2 times brighter than single Cepheids of a similar pulsation period, and have already proven an unmatched observational opportunity to find Cepheids with giant companions \citep{pilecki21}. However, in the context of distance determinations, such systems could bias the results, and therefore are routinely removed from the sample in the process of data cleaning.

On the other hand, binary Cepheids with MS companions (color-coded with navy blue in Figure \ref{fig:percentLMC}), which lie very close to the PLR, are virtually undetectable via photometric methods and therefore cannot be cleaned from the sample. The light contribution from a MS star to the total brightness of individual systems is minuscule, but because they outnumber companions of other evolutionary types, their cumulative effect on the PLR might not be negligible. This effect cannot be assessed from the visual inspection of Figure \ref{fig:percentLMC} alone, therefore we quantify it in Section \ref{sec:DeltaSLZP_bin}.

\subsection{Extension of the synthetic magnitudes}

In \citetalias{karczmarek22} we calculated for all Cepheids and their companions UBVRIJHK magnitudes in the Bessell \& Brett photometric system \citep{bessell90,bessell98} using the online YBC database\footnote{\url{http://stev.oapd.inaf.it/YBC}, accessed January 22, 2022} of stellar bolometric corrections \citep{chen19}. In this work, we extend the list of passbands (using the same online YBC database) to the Vera Rubin Observatory Telescope system magnitudes (ugrizY), and those used by current space missions: Gaia ($G$, $G_\mathrm{BP}$, $G_\mathrm{RP}$), Hubble Space Telescope (F555W, F814W, F160W), and James Webb Space Telescope bands (F070W, F090W, F115W, F150W, F200W, F277W, F356W, F444W).
In addition to the reddening-free Wesenheit index $W_{VI} = I - 1.55 (V-I)$ used in \citetalias{karczmarek22}, we also calculate
\begin{eqnarray}
    W_G &=& G - 1.9 (G_\mathrm{BP} - G_\mathrm{RP}) ~~ \mbox{\citep{ripepi19},} \nonumber \\
    W_{H} &=& \mathrm{F160W} - 0.386 (\mathrm{F555W} - \mathrm{F814W}) ~~ \mbox{\citep{riess21}.} \nonumber    
\end{eqnarray}

We extended the list of magnitudes in anticipation of the vast and various data coming from the Vera Rubin and James Webb Space Telescope, which will also target classical Cepheids. The list of Wesenheit indices was extended due to the recent rise in popularity of the period-Wesenheit and period-Wesenheit-metallicity relations from the Gaia data \citep{lin22,ripepi22}, and due to the ongoing discussion about the precision and accuracy of the Hubble-Lema\^{i}tre constant, $H_0$, the value of which was calculated---among others---using the $W_H$ index of classical Cepheids \citep{riess22}.

\section{The effect of binarity on the slope and the zero point}
\label{sec:DeltaSLZP_bin}

In order to calculate the cumulative effect of binary Cepheids on the PLR, we followed a standard routine of distance determination: (i) cleaning the data from the outliers by applying $3\sigma$ clipping iteratively three times; (ii) least-squares linear fitting to the cleaned data to the Leavitt Law \citep{leavitt1912}:
\begin{equation}
M = sl \times \log P + zp
\end{equation}
with the slope ($sl$) and intercept (zero point, $zp$) as free parameters; (iii) determining the distance from the zero point of the PLR. The values and uncertainties of the slope and zero point were calculated using a bootstrap technique; the least-squares fit was performed on randomly chosen 50\% of the sample (i.e. 5000 systems), the values of the slope and zero point were saved, and the routine was repeated 1000 times, producing normal distributions of the slope and zero point. Finally, a Gaussian fit yielded the mean and standard deviation, which we recognized as a true value and its uncertainty, respectively, of the PLR slope and the zero point. This calculation was performed for the five different adopted binary percentages $f_\mathrm{bin}$ = 0, 25, 50, 75, 100\%, on all synthetic populations in our collection.

None of the environments we simulate has exactly 5000 Cepheids; the number of Cepheids in the LMC and SMC is 4620 and 4915, respectively, with the completeness of nearly 100\% \citep{soszynski15ccep}, and the number of Galactic Cepheids is 3352 with the completeness of about 88\% down to a magnitude $G = 18$ \citep{pietrukowicz21}. However, by keeping our populations equal in size, we assure that statistical errors affecting the samples remain the same, allowing a fair comparison.

\subsection{Shift of the PLR slope}
\label{sec:sl_shift}

\begin{figure}
\centering
\includegraphics[width=\columnwidth]{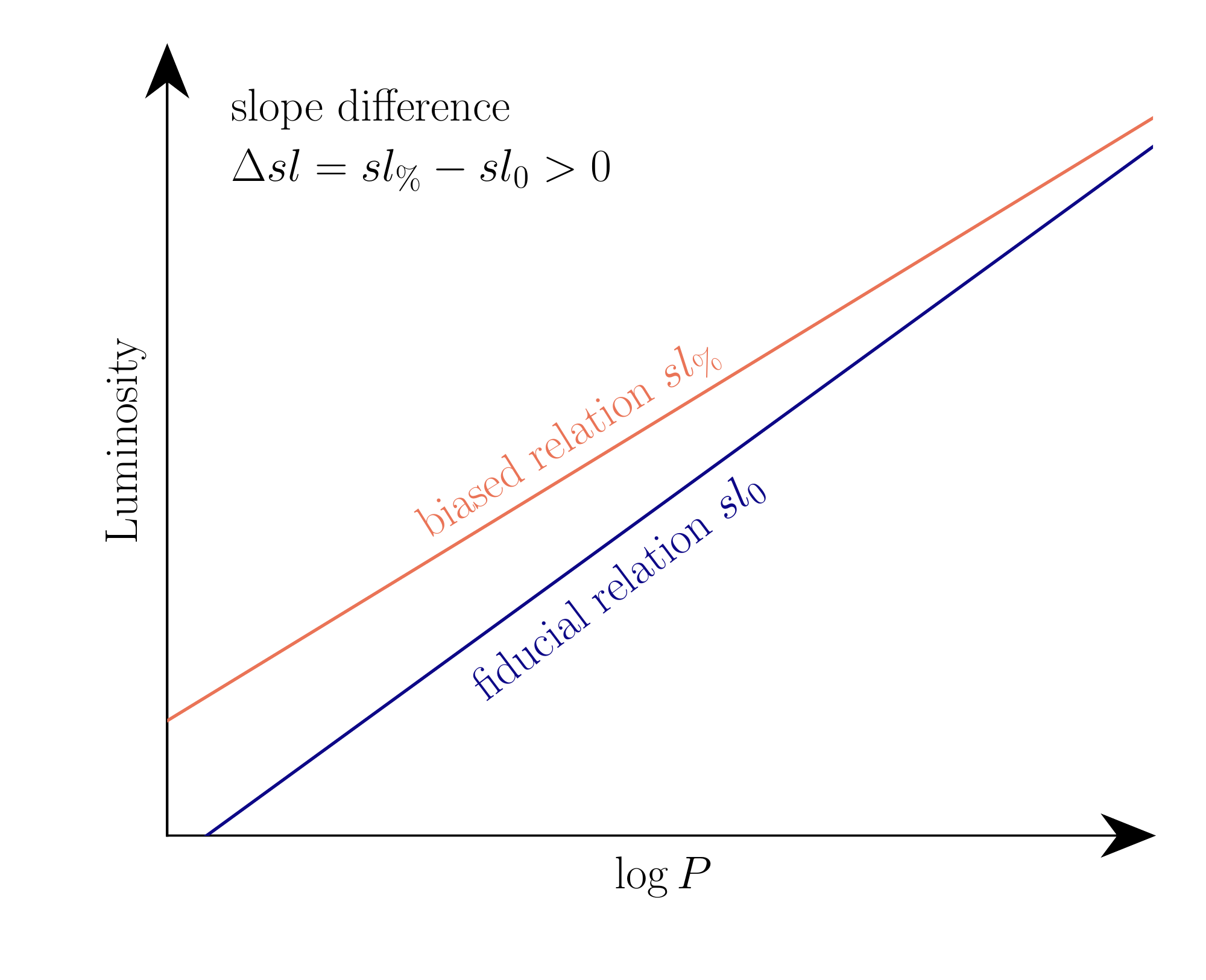}
\caption{Qualitative representation of the effect the companions have on the slope of the period-luminosity relation. \label{fig:sl_sketch}}
\end{figure}

Figure \ref{fig:sl_sketch} presents a qualitative representation of the effect companions have on the slope of the PLRs. Binary Cepheids are expected to flatten the slope of the PLR, because the light of the companion constitutes a bigger portion of the total light for less luminous Cepheids, which are on the low-mass and short-period end of the distribution. High-mass and long-period Cepheids, on the contrary, are more luminous and their light dominates the total light of the system, meaning that the contribution of the companion is considerably reduced. As a result, the slope of the PLR for binary Cepheids is less steep than for single Cepheids, which can be measured as a difference between the two slope values:
\begin{equation}
    \Delta sl = sl_{\%} - sl_0 \,,
\end{equation}
where $sl_0$ is the PLR slope in a pure sample of single Cepheids ($\fbin=0$\%), and $sl_{\%}$ is the PLR slope in a mixed population of single+binary Cepheids, with the binarity percentage in the subscript. The value $\Delta sl$ is therefore expected to be more positive, appearing shallower.

The error on $\Delta sl$ was conservatively taken as the bigger of the two standard deviations: $\sigma_{sl_0}$, $\sigma_{sl_{\%}}$. A typical way of assessing errors, i.e. adding in quadrature, has not been done here since the results are correlated (derived from the same sets of data). 

\begin{figure*}
    \centering
    \includegraphics[width=0.95\textwidth]{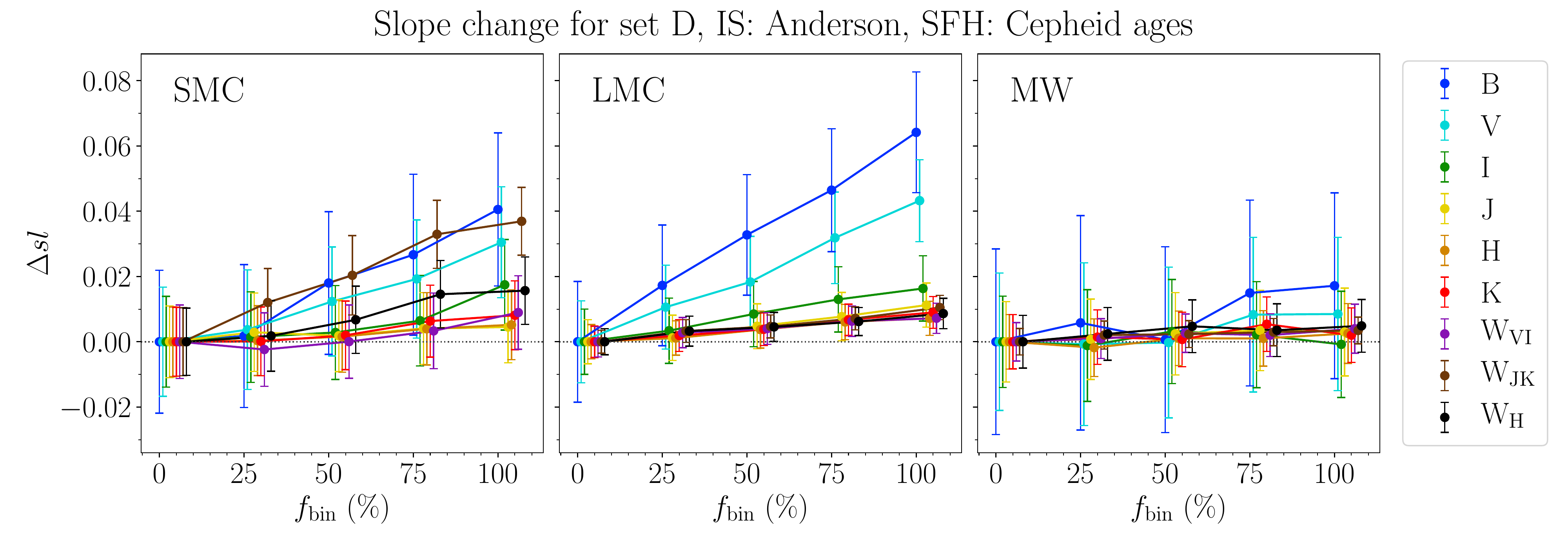}
    \caption{The difference between the slope values of mixed samples of single+binary Cepheids (binarity percentage $\fbin>0$) and a pure sample of single Cepheids ($\fbin=0$). The different colors represent the passbands in the visual and NIR domains, and the Wesenheit indices. Points of different colors are slightly shifted for clarity. The panels show the results for the synthetic population created from set D of the initial parameter distributions, IS prescription of Anderson, and SFH based on the age distribution of the Cepheids, for the three considered metallicity values. The complete figure set (16 images) for all variants of synthetic populations is available in the online journal.}
    \label{fig:DeltaSL}
\end{figure*}

The values of $\Delta sl$ for three variants of synthetic populations, having the metallicities of the SMC, LMC, and MW (with other characteristics fixed to: set D of the initial parameter distributions, the IS prescription of Anderson, and the SFH based on the age distribution of the Cepheids) are presented in Figure \ref{fig:DeltaSL}. Generally, larger binary fractions result in shallower PLR slopes, and larger $\Delta sl$ values, as expected. This effect seems to be more pronounced in the visual domain than in the NIR, however, the large uncertainties on the slope values prevent its accurate determination and make it consistent with zero. In the B-band, whose slope is around $2.0-2.5$ \citepalias{karczmarek22}, the slope change $\Delta sl=0.06$, as presented in Figure \ref{fig:DeltaSL}, translates to a relative error of $2-3$\% ($\Delta sl/sl_0$) and sets the upper limit for the slope error. However, the $B$ band is not used for distance determinations, and more frequently used NIR and Wesenheit magnitudes have a relative error of just 0.3\% or less. Similar results are obtained for other variants of synthetic populations (plots are provided in the online material). Based on this result we state that \emph{the value of the PLR slope remains constant within uncertainties for all binarity percentages and passbands}.

\subsection{Shift of the PLR zero point}

\begin{figure}
\centering
\includegraphics[width=\columnwidth]{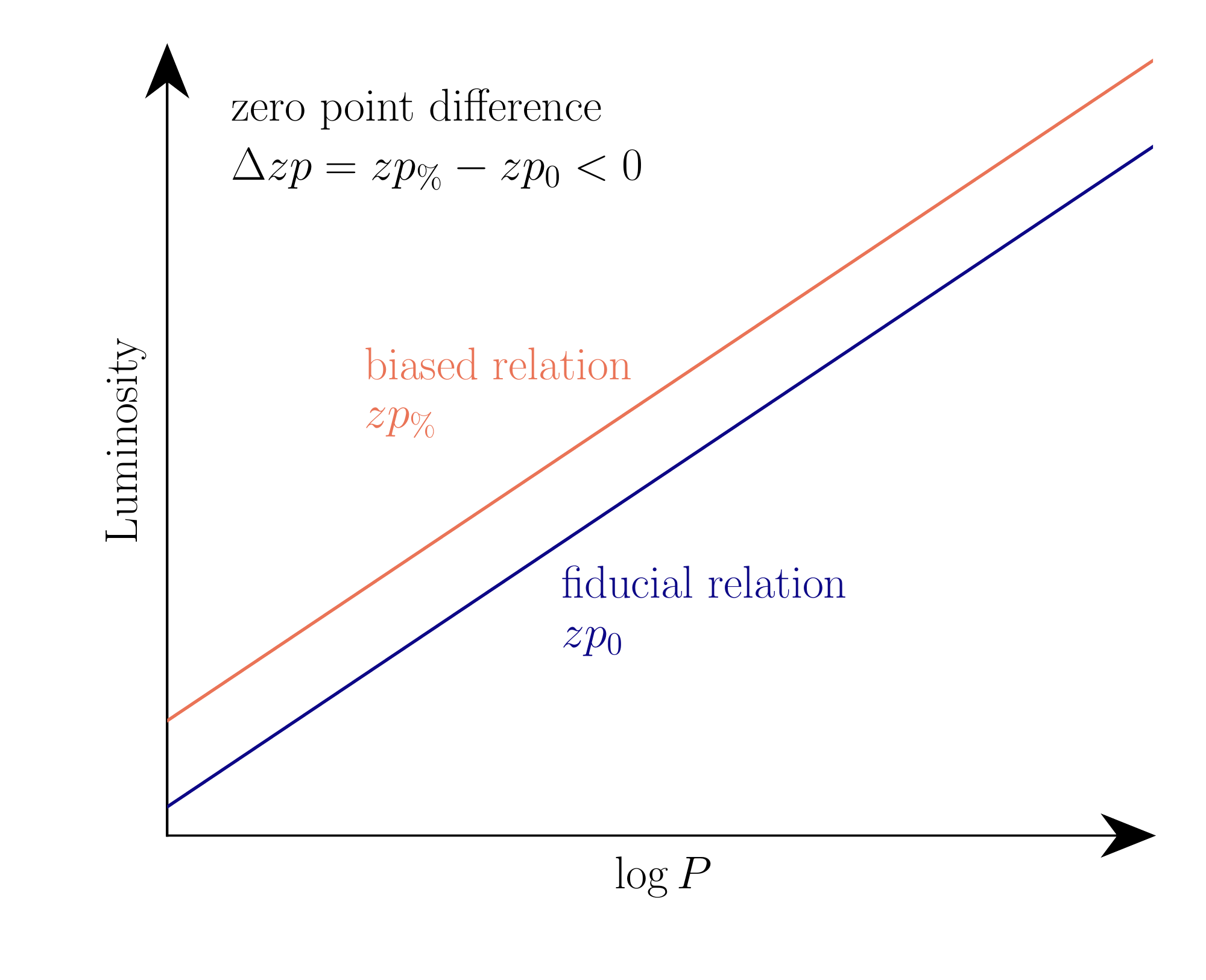}
\caption{Qualitative representation of the effect the companions have on the zero-point of the period-luminosity relation. \label{fig:zp_sketch}}
\end{figure}

We now consider the effect of companions on the zero point of the PLR, when the slope is fixed (Figure \ref{fig:zp_sketch}). Binary Cepheids are expected to shift the zero point of the PLR toward brighter magnitudes. The zero point of the PLR for binary Cepheids is therefore brighter than for single Cepheids, which can be calculated as a difference between the two zero point values:
\begin{equation}
    \label{eq:DeltaZP}
    \Delta zp = zp_{\%} - zp_0 \,,
\end{equation}
where $zp_0$ is the PLR zero point in a pure sample of single Cepheids, and $zp_{\%}$ is the PLR zero point in a mixed population of single+binary Cepheids, with the binarity percentage in the subscript. The value $\Delta zp$ is therefore expected to be more negative for larger upward shifts.

\begin{figure*}
    \centering
    \includegraphics[width=0.95\textwidth]{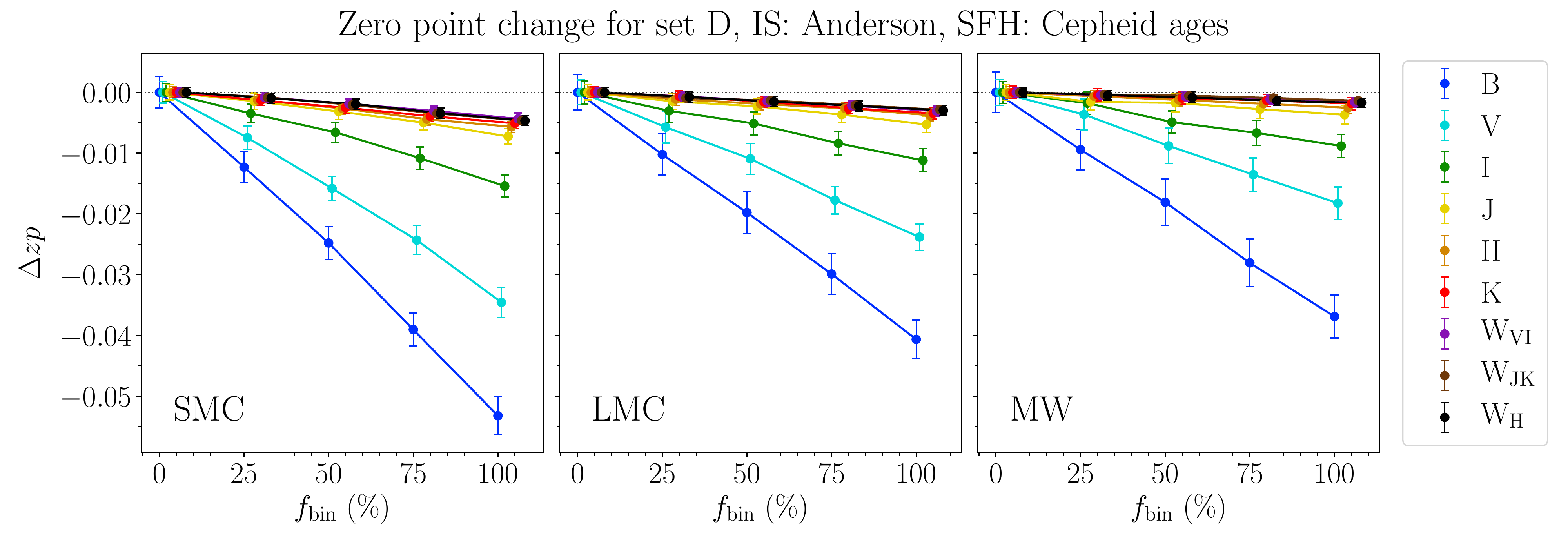}
    \caption{The difference between the zero point values of mixed samples of single+binary Cepheids (binarity percentage $\fbin>0$) and a pure sample of single Cepheids ($\fbin=0$). The different colors represent the passbands in the visual and NIR domains, and the Wesenheit indices. Points of different colors are slightly shifted for clarity. The panels show the results for the synthetic population created from set D of the initial parameter distributions, IS prescription of Anderson, and SFH based on the age distribution of the Cepheids, for the three considered metallicity values. The complete figure set (16 images) for all variants of synthetic populations is available in the online journal.}
    \label{fig:DeltaZP}
\end{figure*}

In order to calculate $\Delta zp$, we followed the same procedure for generic distance determination as described above, but this time we fitted only the zero points while the slopes were fixed to the value $sl_0$ that we had found for a pure sample of single Cepheids in Section \ref{sec:sl_shift}. Fixing the slope on $sl_0$ is arbitrary, and choosing another value, e.g. $sl_{50\%}$, does not affect the results. The PLR zero points, determined this way, for the mixed samples of single+binary Cepheids ($zp_{\%}$) and for the pure sample of single Cepheids ($zp_0$), were then used to calculate $\Delta zp$ following Equation (\ref{eq:DeltaZP}). The error on $\Delta zp$ was calculated in the same way as the error on $\Delta sl$.

The values of $\Delta zp$ for three variants of synthetic populations, having the metallicities of the SMC, LMC, and MW (with other characteristics fixed to: set D of the initial parameter distributions, the IS prescription of Anderson, and the SFH based on the age distribution of the Cepheids) are presented in Figure \ref{fig:DeltaZP}. As expected, the bigger the binarity percentage, the larger the upward shift, and the smaller the $\Delta zp$ values. $\Delta zp$ scales linearly with the binarity fraction, and is more pronounced in the visual domain than in the NIR, reaching $-0.05$~mag in the $V$-band, as compared to $-0.01$~mag in the $K$-band, for $\fbin= 100\%$.
Similar results are obtained for all other variants of synthetic populations (the complete set of plots is provided as online material).

\section{The effect of binarity on the distance modulus}
\label{sec:DeltaDM_bin}

The above calculation of the shift of the zero point has been achieved by varying the number of binary Cepheids within the simulated isolated environments of the SMC, LMC, and MW (left, middle, and right panels in Figure \ref{fig:DeltaZP}, respectively). However, extragalactic distances are usually calculated as differences of zero points between two galaxies. In order to properly assess the effect of binarity on the extragalactic distance scale, one needs to consider the shift of the PLR zero point not within one, but between two galaxies, which have different metallicities, and possibly different percentages of Cepheid binaries.

\begin{figure}
    \centering
    \includegraphics[width=\columnwidth]{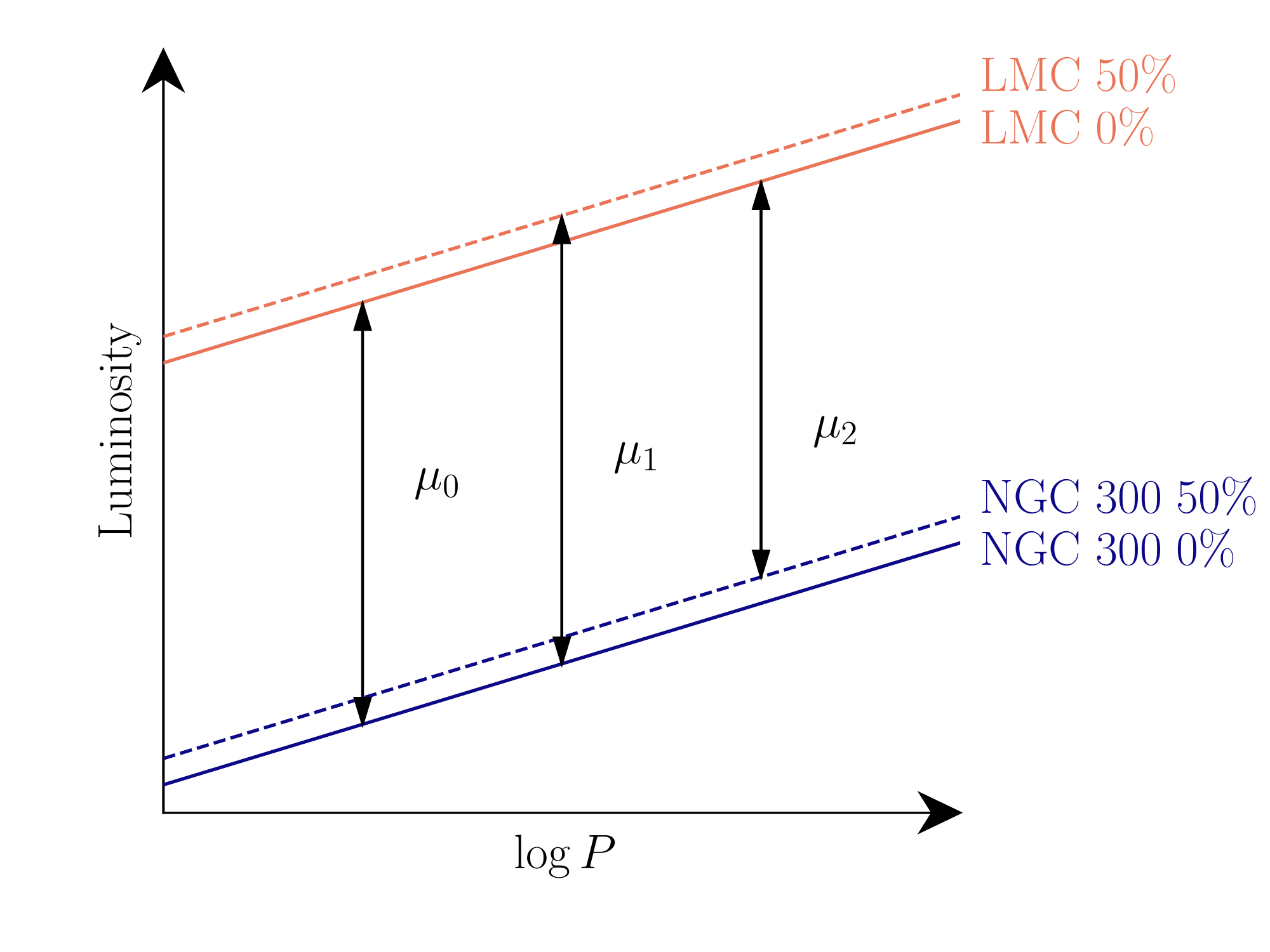}
    \caption{Visualization of three distance moduli ($\mu$) differences between the Cepheid PLRs in the reference (orange, LMC) and target (navy blue, NGC~300) galaxies. $\mu_0$ is the distance modulus between the PLR zero points constructed with single Cepheids only (solid lines).  $\mu_1$ and  $\mu_2$ show how distance moduli would change if the PLR zero points were constructed with 50\% of binary Cepheids (dashed line).}
    \label{fig:dm_sketch}
\end{figure}

In Figure \ref{fig:dm_sketch} we visualize the distance modulus differences between two exemplary galaxies of different metallicities (LMC and NGC~300) and unknown percentage of binary Cepheids (for the sake of argument we propose two values: 0 and 50\%). When calculating the distance modulus between a target and reference galaxy, $\mu = zp^\mathrm{tar} - zp^\mathrm{ref}$, with an unknown percentage of binary Cepheids, we take into account three possibilities: (i) the reference (LMC) and target (NGC~300) galaxies have only single Cepheids, resulting in the distance modulus $\mu_0$; (ii) the reference galaxy has a larger binarity percentage than the target galaxy, resulting in $\mu_1 > \mu_0$; (iii) the reference galaxy has a smaller binarity percentage than the target galaxy, resulting in $\mu_2 < \mu_0$.

The shift in the distance modulus, $\Delta \mu$, is the difference between the distance modulus from the reference to the target galaxy, relative to $\mu_0$:
\begin{equation}
	\label{eq:DeltaDM}
    \Delta \mu = \mu - \mu_0 \,.
\end{equation}
Since the binarity percentages for the reference and target galaxies do not have to be equal, $\Delta \mu$ can be either positive or negative. 

In order to examine the dependence of $\Delta \mu$ on the binarity percentage in the target and reference galaxies, we follow the same procedure for generic distance determination as described in Section \ref{sec:DeltaSLZP_bin}, but this time we assume the PLR slope is the same for all binary percentages and for all metallicities. This slope value is fixed to the LMC slope and $\fbin=0\%$.

\begin{figure}
    \centering
    \includegraphics[width=\columnwidth]{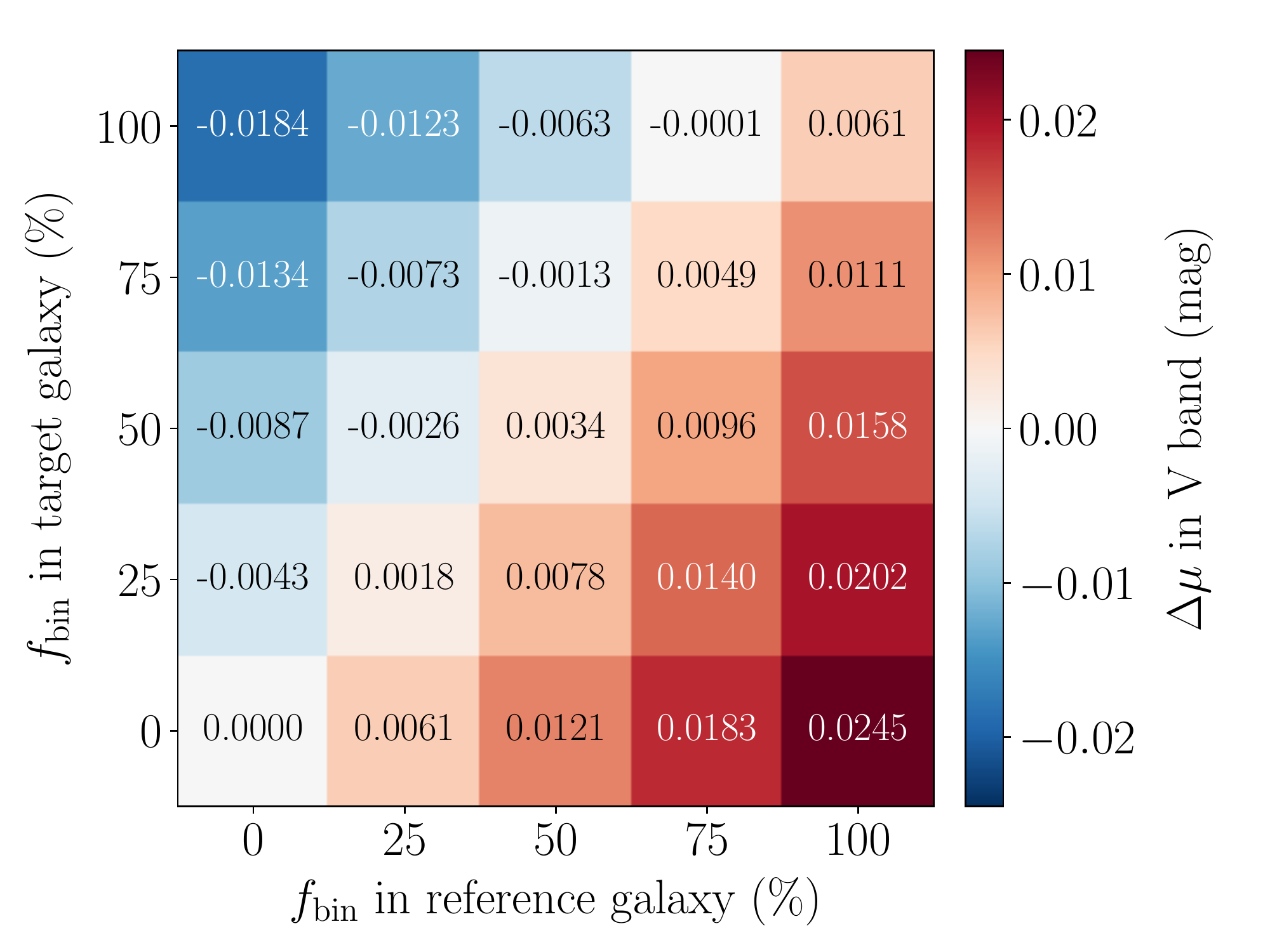}\\
    \includegraphics[width=\columnwidth]{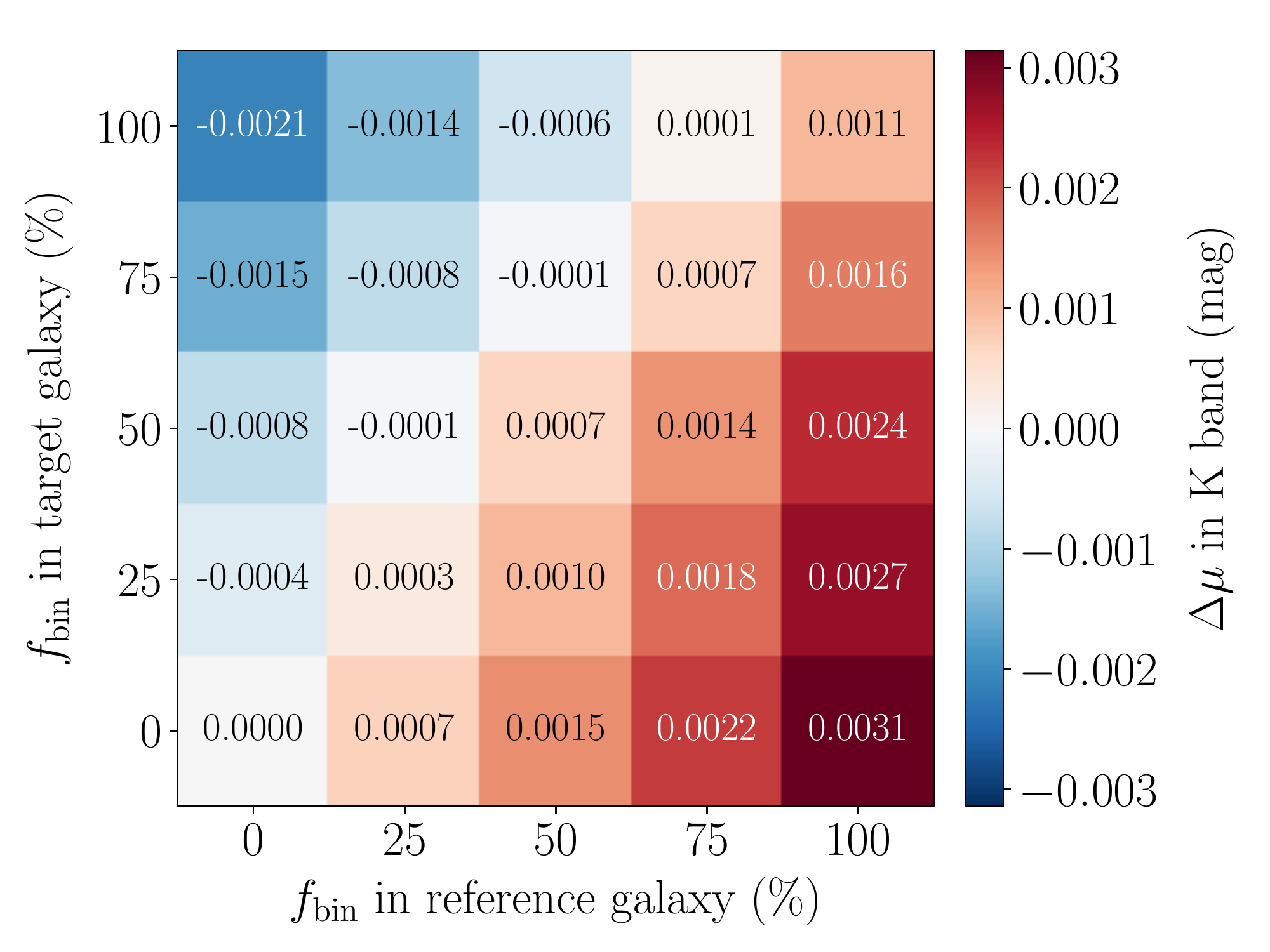}
    \caption{$V$- and $K$-band grids of possible values of distance modulus shifts, $\Delta \mu$, as a function of the binarity percentage in the reference ($Z=0.008$) and target ($Z=0.02$) galaxies. The result is based on the synthetic population variant: set D of the initial parameter distributions, IS prescription of Anderson, and SFH based on the age distribution of the Cepheids.}
    \label{fig:DMshift}
\end{figure}

All possible $\Delta \mu$ values as a function of the binarity percentages of reference and target galaxies are presented in Figure \ref{fig:DMshift} in a form of a matrix. As before, the analyzed synthetic population variant is: set D of the initial parameter distributions, the IS prescription of Anderson, the SFH based on the age distribution of the Cepheids, and the metallicities are $Z=0.008$ and 0.02 for the reference and target galaxy, respectively. Figure \ref{fig:DMshift} shows the results for the $V$ and $K$ band; matrices for other bands and variants are available online (see Section DATA AVAILABILITY).

In each matrix, the bottom left cell represents the baseline value, explicitly $\Delta \mu = \mu_0 - \mu_0 = 0$. In other words, in this cell the distance modulus between two galaxies (both having $\fbin=0\%$) is calculated and subtracted from itself, resulting in a zero value. By contrast, in the bottom middle cell the value of distance modulus difference between the reference ($\fbin = 50\%$) and target ($\fbin = 0\%$) galaxies has been calculated and reduced by the baseline value; this cell corresponds to the value of $\mu_1-\mu_0$ from Figure \ref{fig:dm_sketch}.

Two most extreme values in the top and bottom cells on the antidiagonal ($\searrow$), i.e. $-0.0184$ and 0.0245 ($V$-band), and $-0.0021$ and 0.0031 ($K$-band), show the \emph{maximum systematic error} of the distance modulus, i.e., the maximum value by which the distance modulus is shifted if the ``assumed'' binarity percentage is 0\% while the ``true'' one is 100\%, and vice versa. On the other hand, the diagonal going from the lower left corner to the upper right corner ($\nearrow$) shows the smallest shifts in the distance modulus for the cases of equal or almost equal binarity percentage in the target and reference galaxies. This result means that \emph{the effect of binarity on the extragalactic distance scale is virtually null if the galaxies have the same binary Cepheid fractions.}

If the target and reference galaxies were identical\footnote{Two galaxies would be identical if they had the same chemical composition, SFH, proportions of Cepheids on 1st, 2nd and 3rd crossing, and Cepheid companions of the same distribution of physical characteristics (masses and radii).}, the matrices would be symmetrical, with the diagonal ($\nearrow$) filled with zeros.
However, in more realistic cases the two galaxies have different metallicities, which affects the shape and location of the IS, and physical characteristics (masses and radii) of Cepheid companions. Different SFHs affect the proportions of Cepheids on the 1st, 2nd, and 3rd IS crossings. This further contributes to the dissimilarities between the galaxies, and by extension, to the asymmetries of matrices in Figure \ref{fig:DMshift}.

\subsection{Differences between the variants}
\label{sec:differences}

In order to describe the effect of the aforementioned differences on $\Delta \mu$, we compare the results coming from different variants of synthetic populations. For that, we simplify the matrices from Figure \ref{fig:DMshift} by taking only the smallest value (top left cell) and assuming it is the maximum (in terms of absolute value) systematic error on the distance modulus, i.e. max\,$|\Delta \mu|$. The largest value (bottom right cell) represents the maximum error associated with the choice of a reference galaxy, which throughout our study stays fixed to the LMC, and therefore is of lesser relevance. 

Combinations of two choices of the IS (parallel and Anderson's, `p' and `A' for short) and two choices of SFH (uniform and based on the age distribution of the Cepheids, `u' and `C' for short) result in four variants (AC, pC, Au, pu) for every metallicity and set of the initial parameter distributions. The differences between the variants are grouped by: (i) choice of IS and SFH, (ii) metallicity, (iii) set of the initial parameter distributions; and presented as a function of wavelength. Mind that the results represent the maximum systematic error, and most likely are much smaller if the target and reference galaxies have similar fractions of binary Cepheids.

\begin{figure}
    \centering
    \includegraphics[width=\columnwidth]{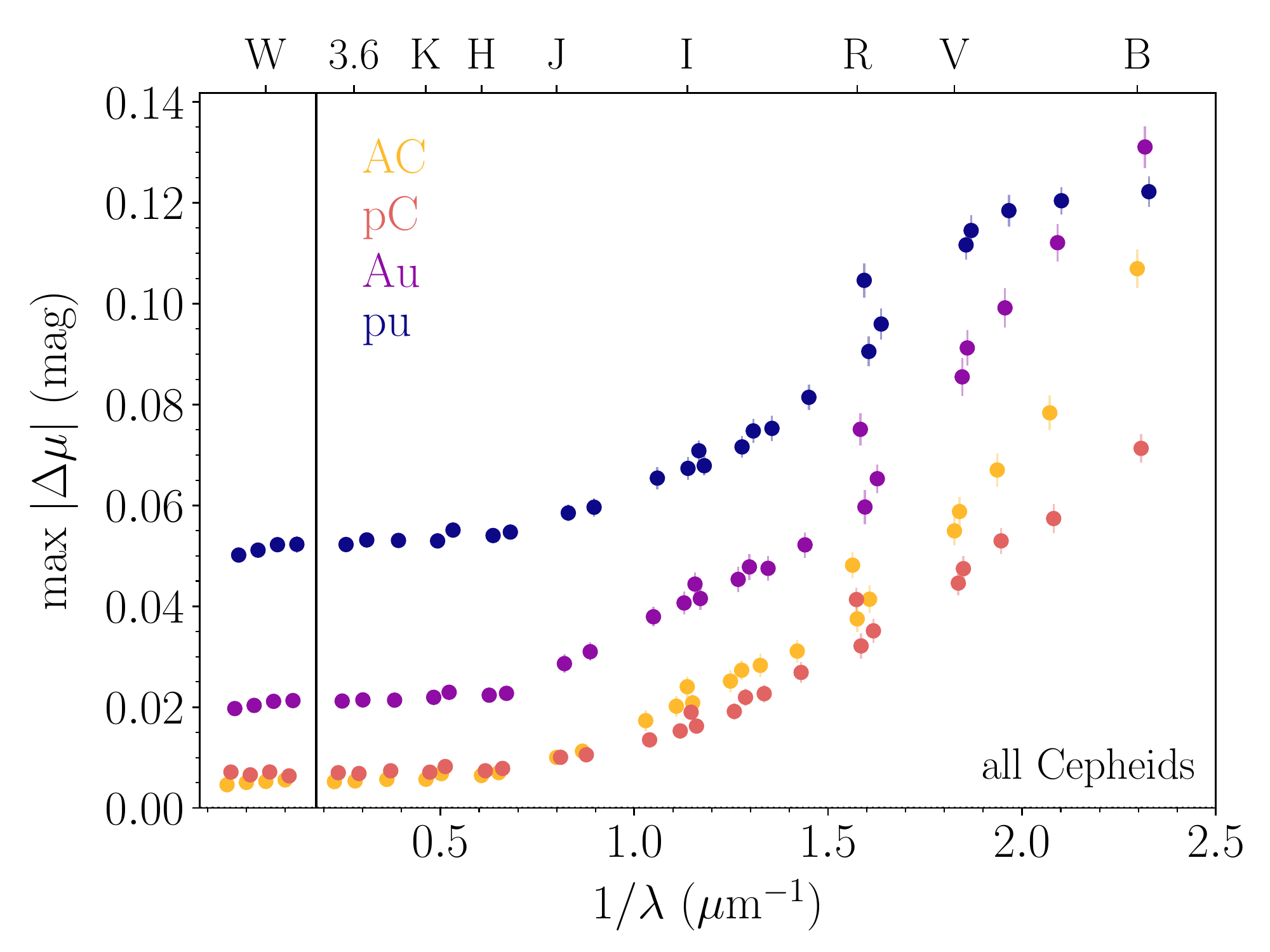}\\
    \includegraphics[width=\columnwidth]{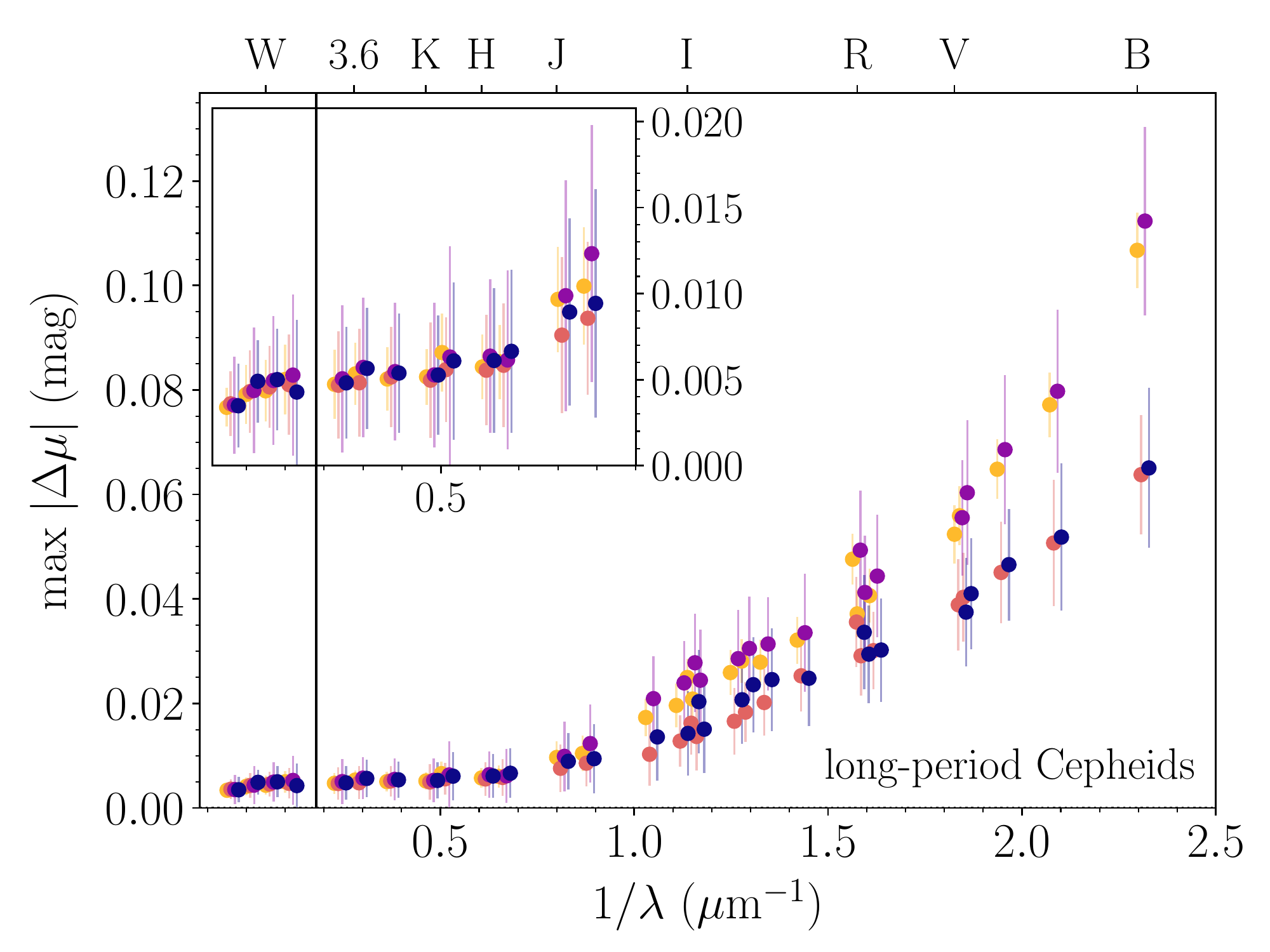}
    \caption{The maximum absolute value of $\Delta \mu$ as a function of wavelengths of the considered bands, as well as for the Wesenheit indices, for the complete sample (top) and for long-period Cepheids only (bottom). The colors correspond to four variants of synthetic populations of binary Cepheids: AC, pC, Au, pu (see text for detail); they all have $Z=0.02$ and come from set A of the initial parameter distributions. The points corresponding to the same band or index are shifted slightly for clarity. On the left side from the vertical black solid line are Wesenheit indices in order from left to right: $W_{H}$, $W_{JK}$, $W_{VI}$, $W_G$. The inset in the lower panel zooms in on the differences between the Wesenheit indices and the longest-wavelength bands considered here. Larger uncertainties in the lower panel are due to the smaller long-period Cepheid sample (compared to the total sample on the top panel).}
    \label{fig:DeltaDM_variants}
\end{figure}

\paragraph{Choice of the IS and SFH}
Figure \ref{fig:DeltaDM_variants} shows four variants of synthetic populations (AC, pC, Au, pu) for set A of the initial parameter distributions and $Z=0.02$. We exceptionally present them---instead of usually used set D---to illustrate the most extreme cases, as set A yields the largest systematic error on the distance modulus among all our variants.

All four variants of set A, presented in Figure \ref{fig:DeltaDM_variants}, show the same trend that max\,$|\Delta \mu|$ decreases with increasing wavelength and reaches the smallest values in the near- and mid-infrared, along the Wesenheit indices (in the figure in order from left to right: $W_{H}$, $W_{JK}$, $W_{VI}$, $W_G$). However, variants Au and pu (i.e. with uniform SFH) show conspicuous and unrealistically high values of max\,$|\Delta \mu|$, even in the infrared domain. This is due to the very high percentage of IS1 Cepheids in the samples (cf. Figure 4 in \citetalias{karczmarek22}), whose brightness is dominated by the extra light from the companion\footnote{%
High percentages of IS1 Cepheids in metal-rich samples with uniform SFH are artifacts, and arise from the caveat of theoretical models of stellar evolution, which fail to render blue loops for metal-rich stars of masses below $\sim 3.5 \msun$. As a result, such stars traverse the IS only once in the Hertzsprung Gap, which creates an deficiency of IS2+3 crossers, and therefore and overabundance of IS1 Cepheids (see more in Section 3.2 of paper I). This issue disappears for high-mass metal-rich stars and metal-poor stars of all masses, whose evolutionary tracks show pronounced blue loops that overlap with the IS.}.
Indeed, for a metal-poor environment ($Z=0.004$) with much smaller ratio of IS1 to IS2+3 Cepheids, trends marked by the Au and pu variants overlap with the AC and pC trends. In fact, if we consider only long-period Cepheids ($P_\mathrm{pul}>10$~d, bottom panel of Figure \ref{fig:DeltaDM_variants}), the shape of the IS becomes the differentiating factor in the visual domain (smaller max\,$|\Delta \mu|$ are expected from variants with parallel IS) and in the infrared all four variants overlap again. Larger error bars stem from the smaller sample size, consisting of only long-period Cepheids as opposed to the full-sized sample of ``all-period" Cepheids.
While considering only long-period Cepheids increases the uncertainty, it does not change the results for the AC and pC variants.

The analysis above further affirms that variants with uniform SFH (Au and pu) produce unrealistic results due to the unrealistically high ratio of IS1 to IS2+3 Cepheids in the sample \citepalias[already mentioned in][]{karczmarek22}. Therefore, further analysis will focus only on variants created from the SFH based on the age distribution of the Cepheids (AC and pC).

\paragraph{Metallicity}
The values of max\,$|\Delta \mu|$ are by definition metallicity-reduced, in the sense that the effect of metallicity on the zero point of the PLR \citep[described as $\gamma$ term, e.g.][]{gieren18,breuval21,breuval22}, has been removed at the stage of subtracting the baseline distance modulus $\mu_0$ (recall equation \ref{eq:DeltaDM}). However, the results are not entirely metallicity-independent, because the metallicity affects not only the brightness of Cepheids (and therefore the PRL zero point via the $\gamma$ term) but also their companions.

\begin{figure}
    \centering
    \includegraphics[width=\columnwidth]{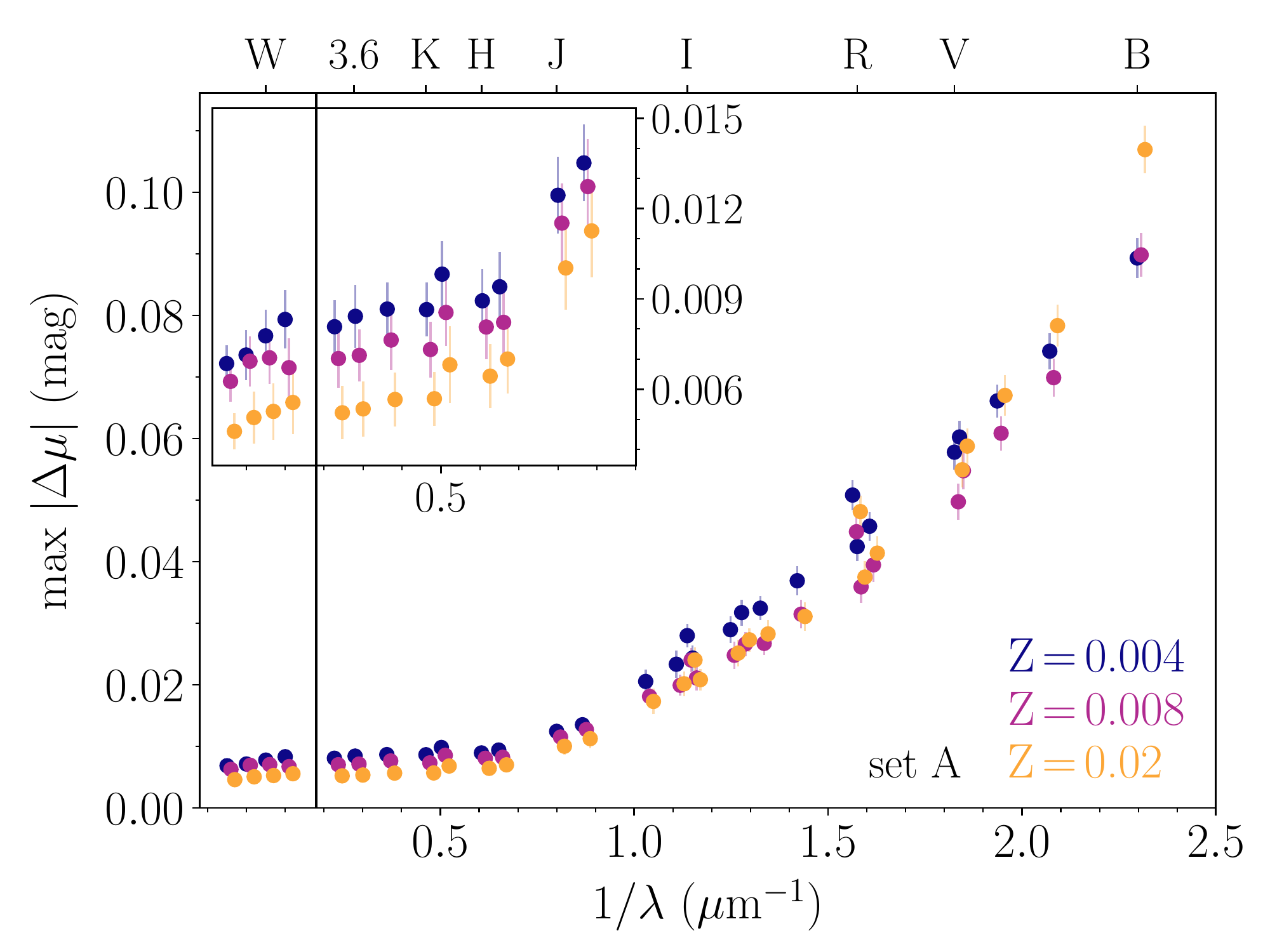}\\
    \includegraphics[width=\columnwidth]{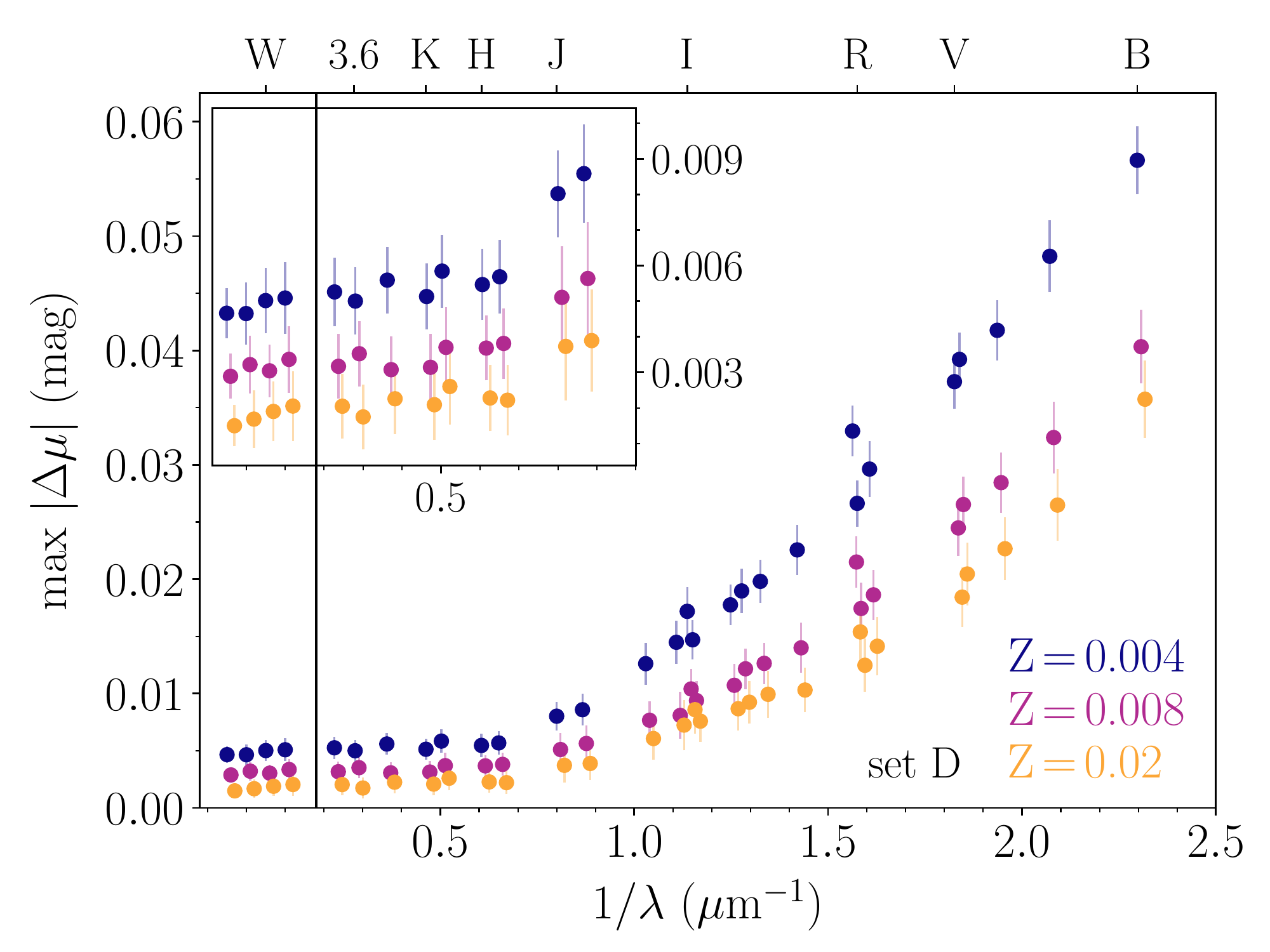}
    \caption{The maximum absolute value of $\Delta \mu$ as a function of wavelengths of the considered bands, as well as for the Wesenheit indices, for the complete Cepheid sample of three metallicities $Z=0.004$, 0.008, and 0.02 (color-coded). The presented variants correspond to set A (top) and D (bottom) of the initial parameter distributions, Anderson's prescription of the IS and the SFH based on the age distribution of the Cepheids. The points corresponding to the same band or index are shifted slightly for clarity. On the left side from the vertical black solid line are Wesenheit indices in order from left to right: $W_{H}$, $W_{JK}$, $W_{VI}$, $W_G$. The insets zoom in on the differences between the Wesenheit indices and the longest-wavelength bands considered here.}
    \label{fig:DeltaDM_Z}
\end{figure}

Specifically, in metal-poor environments Cepheids evolve along more pronounced blue loops on the Hertzsprung-Russell diagram at lower luminosities than in the metal-rich ones. This means that the population of metal-poor Cepheids is systematically older than the metal-rich one, and therefore is more likely to have red giant companions, because the companions had had more time to evolve (Richard Anderson, private communication). Indeed, Figure \ref{fig:DeltaDM_Z} shows that max\,$|\Delta \mu|$ are larger for $Z=0.004$ than for 0.02 in the infrared domain, which confirms that the excess light comes from red companions. This trend reverses, however, for shorter wavelengths, and in the visual domain max\,$|\Delta \mu|$ is the largest for $Z=0.02$. The flip in the trend occurs for sets A and B of the initial parameter distributions, while for sets C and D the trend remains the same, i.e., max\,$|\Delta \mu|$ is larger for lower metallicities at all wavelengths. The factor responsible for this dichotomy is the distribution of the initial mass ratio; the effect of the initial parameter distributions is described below. The shape of the IS (either metallicity-independent parallel IS or metallicity-dependent wedge-like IS) has no significant effect on the results. 

\begin{figure}
    \centering
    \includegraphics[width=\columnwidth]{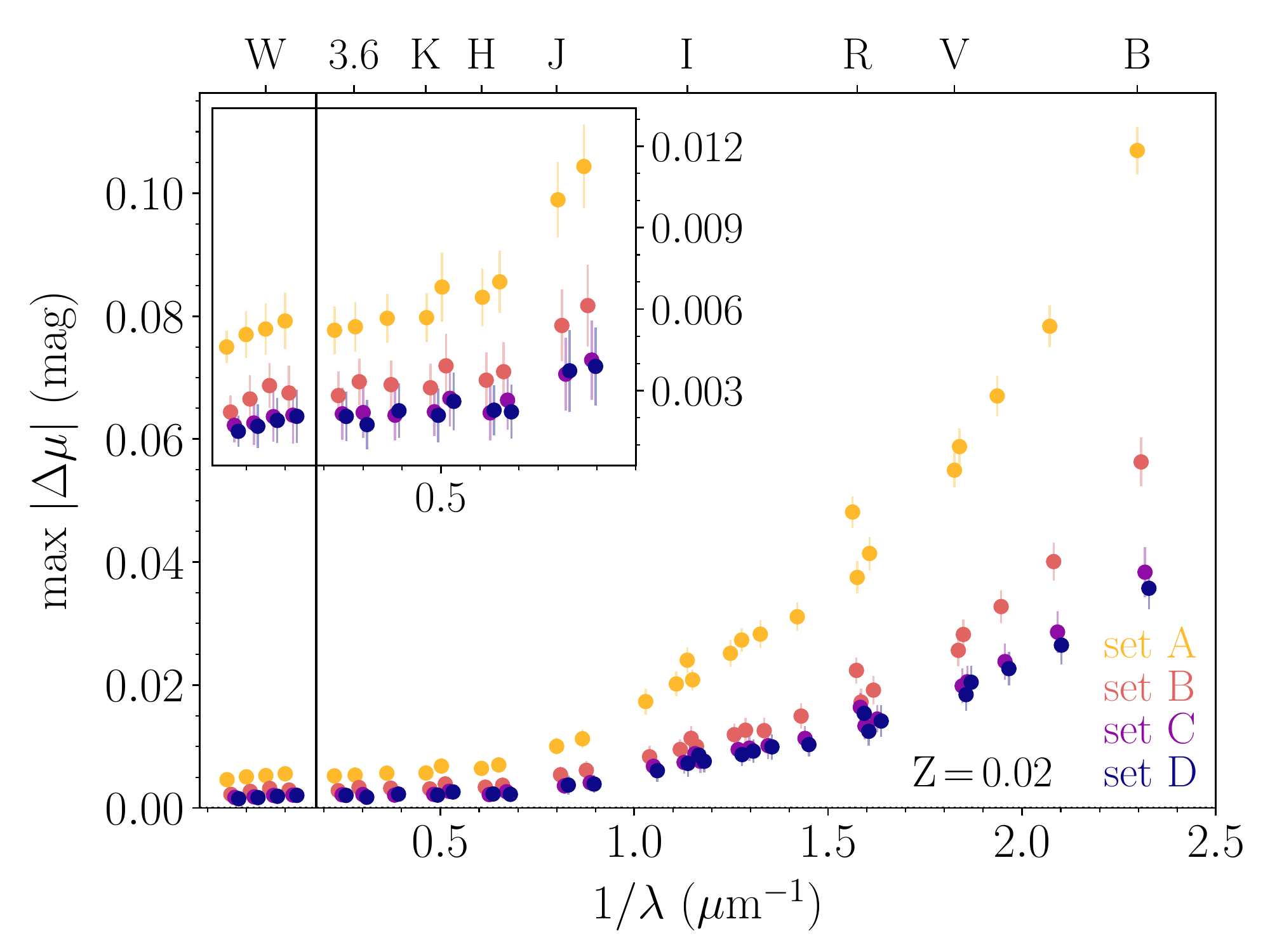}\\
    \includegraphics[width=\columnwidth]{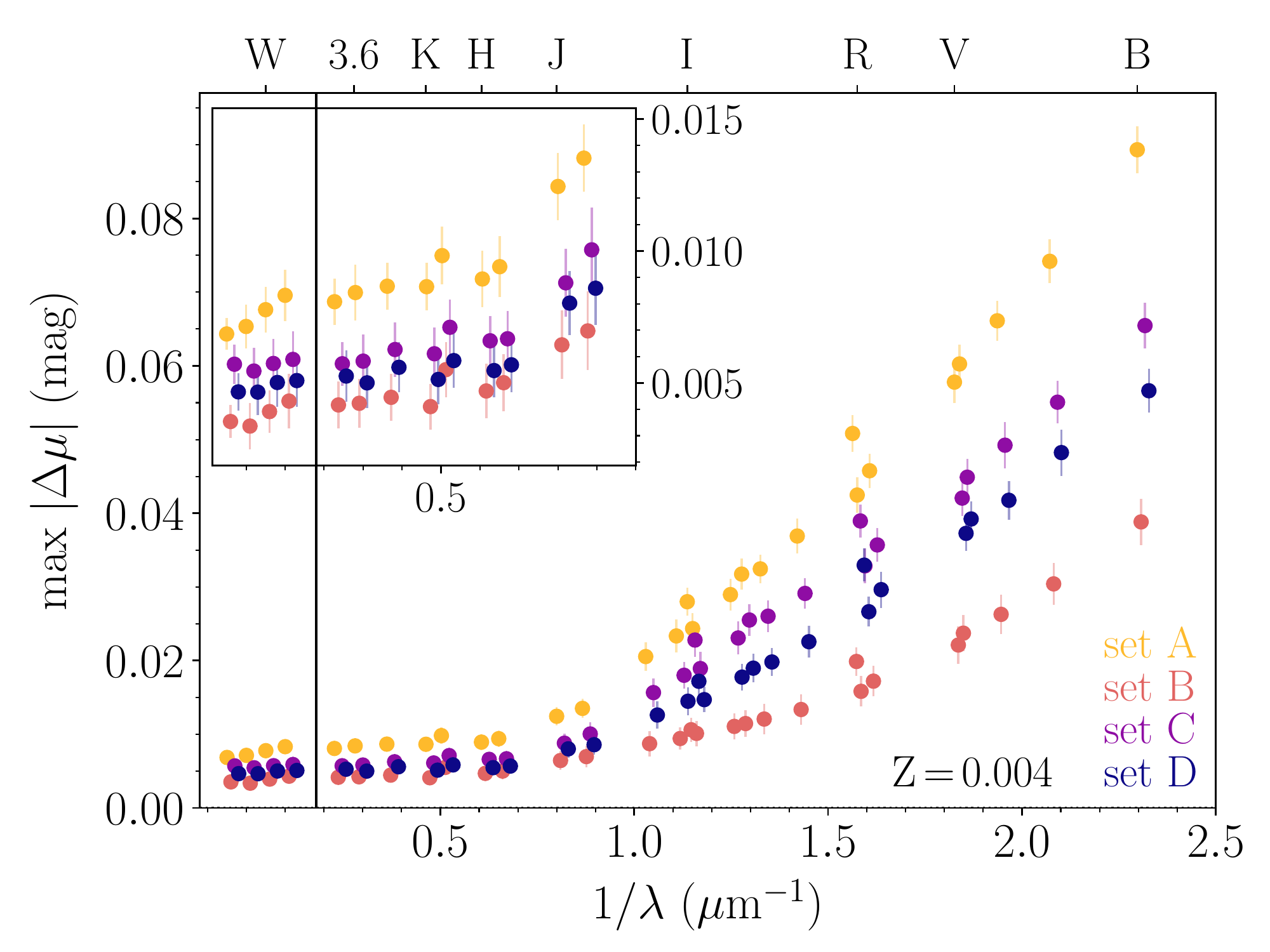}
    \caption{The maximum absolute value of $\Delta \mu$ as a function of wavelengths of the considered bands, as well as for the Wesenheit indices, for the complete Cepheid sample coming from four sets of the initial parameter distributions (color-coded). Presented variants have metallicities $Z=0.02$ (top) and 0.004 (bottom), Anderson's prescription of the IS, and the SFH based on the age distribution of the Cepheids. The points corresponding to the same band or index are shifted slightly for clarity. On the left side from the vertical black solid line are Wesenheit indices in order from left to right: $W_{H}$, $W_{JK}$, $W_{VI}$, $W_G$. The insets zoom in on the differences between the Wesenheit indices and the longest-wavelength bands considered here.}
    \label{fig:DeltaDM_sets}
\end{figure}

\paragraph{Sets of the initial parameter distributions}
Out of four sets of the initial parameter distributions (A, B, C, D; for details see Table 1 and Figure A1 in \citetalias{karczmarek22}), set A gives the largest max\,$|\Delta \mu|$ for every metallicity and variant of IS and SFH, which is around 0.06~mag in the visual (for all metallicities) and 0.006-0.009~mag in the infrared (larger for lower metallicity), as shown in Figure \ref{fig:DeltaDM_sets}. Set B shows similar behavior, though max\,$|\Delta \mu|$ is systematically smaller than for set A, and is around 0.02~mag in visual (for all metallicities) and 0.003-0.005~mag in infrared (larger for lower metallicity). Sets C and D show larger discrepancies with metallicity; their max\,$|\Delta \mu|$ values reside between the trends marked with A and B values for $Z=0.004$ (Figure \ref{fig:DeltaDM_sets}, bottom), but drop below set B for $Z=0.02$ (Figure \ref{fig:DeltaDM_sets}, top), having the lowest max\,$|\Delta \mu| \approx 0.002$~mag in the infrared domain out of all variants of the synthetic populations. This interplay between the metallicity and sets of the initial parameter distributions has already been noticed in the previous paragraph and here it is investigated in more detail.

The crucial difference between the sets of the initial parameter distributions lies in the distributions of the mass ratio, $q=M_\mathrm{B}/M_\mathrm{A}$. This distribution is uniform (set A), log-normal (set B), or described by a decreasing power-law combined with a peak at $q=1$ (sets C and D). In other words, in set A Cepheids have companions with a uniformly distributed mass between the lower ($0.08~\msun$) and upper limit (the mass of the Cepheid), while in set B low-mass companions (with $q \approx 0.3$) dominate. In sets C and D even lower-mass companions prevail ($q \approx 0.1$), but companions with similar masses to the Cepheids are also present in the sample to a greater (set C) or lesser extent (set D).

Cepheids from set A have companions of uniformly distributed mass, which means that neither high-mass evolved red giants nor low-mass and faint MS stars dominate the sample. As a result, in Figure \ref{fig:DeltaDM_sets} medium-mass hot MS stars contribute the most to the value of $|\Delta \mu|$ at shorter wavelengths, while high-mass evolved red giants---at longer wavelengths. Unlike set A, set B consists of Cepheid companions that are mostly low- to medium-mass MS stars, which explains systematically smaller values of max\,$|\Delta \mu|$ as compared to set A. Finally, sets C and D have more high-mass than medium-mass stars, and as a result the values of max\,$|\Delta \mu|$ in Figure \ref{fig:DeltaDM_sets} lie between trends marked by sets A and B in the metal-poor environment (because high-mass companions had time to evolve into red giants; bottom panel) and below the set-B trend in metal-rich environments (top panel).

\section{The effect of binarity on the Hubble-Lema\^{i}tre constant}
\label{sec:H0}

The Hubble-Lema\^{i}tre constant, $H_0$, has been determined by the SH0ES Program \citep[][and references therein]{riess22} in three steps, and anchored in three galaxies: MW, LMC, and NGC~4258. These galaxies, apart from hosting classical Cepheids, have distances determined from purely geometric methods (Gaia parallaxes in the MW, detached eclipsing binaries in the LMC, and water masers in the nucleus of NGC~4258), which allows calibrating the zero point of the Cepheid PLR (step 1). Next, the calibrated PLR is used to determine distances to galaxies that host both Cepheids and type Ia supernovae (SNeIa), and by doing so, to calibrate SNeIa as distance indicators (step 2). Finally, the SNeIa at high redshifts, whose distances have been calculated in step 2, yield $H_0$ (step 3).

The calibration of the zero point of the Cepheid PLR (step 1) has been repeatedly refined by including factors like metallicity \citep{breuval22}, crowding, or reddening \citep{riess22}. However, the effect of binarity has not been taken into account yet. From Figure \ref{fig:DeltaZP} we expect that the zero point of the Cepheid PLR in the LMC should be shifted by about 0.02 and 0.003~mag in the $V$ and $K$ bands, respectively, if the binarity percentage is $\fbin=100\%$, and half of these values if $\fbin=50\%$. In the MW, the shifts are 0.016 and 0.002~mag in the $V$ and $K$ bands, respectively ($\fbin=100\%$). However, none of these values matter for the calibration of the PLR zero point at step 1. In principle, the calibration serves to establish a dependence between the geometrical distance and the value of the PLR zero point, thus the absolute value of the zero point is irrelevant as long as this very calibration is preserved in all Cepheid populations. 

If we were able to exclude binary Cepheids or eliminate the extra light of the companions in the three anchor galaxies, and calibrate the zero point of the PLR based only on pure samples of single Cepheids, then such a calibration would prove useful only if Cepheids in the SNeIa host galaxies were also all single or if the adjustment for the extra light from the companions was introduced to the calibration. Since the existing census of binary Cepheids suffers from tremendous completeness biases, which render it impossible to remove the companions, the only viable alternative is to include binary Cepheids into the PLR calibration.

Binary Cepheids do not compromise the accuracy of the PLR calibration at step 1, but they might decrease the precision of this calibration if the percentages of binary Cepheids in the three anchor galaxies vary considerably. This might cause a scatter in the PLR calibration averaged from the three anchors. However, this statistical error at its maximum ($\fbin=100\%$) is expected to be around 0.004 and 0.001~mag in the $V$ and $K$ bands, respectively, and most probably it is even smaller.

Binary Cepheids play a much bigger role at step 2 of the determination of $H_0$, when the already established PLR calibration is used to determine distances to SNeIa host galaxies. The premise is that Cepheids in anchor and SNeIa host galaxies follow the same PLR, and any significant differences in their brightnesses, attributed to crowding, reddening, or different metallicities, are already taken into account in the $H_0$ determination. Our study shows that different binarity percentages in anchor and SNeIa host galaxies cause a shift in the distance modulus between the two galaxies, which can be either positive (galaxies appear to be farther away from each other) or negative (galaxies appear to be closer to each other). The maximum value of the shift (if the values of $\fbin$ in reference and target galaxies are 0 and 100\%, or vice versa) is $|\Delta \mu| \approx 0.008$~mag in $W_H$, but is substantially minimized if the values of $\fbin$ are similar in the two investigated galaxies. 

Currently, step 2 of the $H_0$ determination involves 37 SNeIa host galaxies \citep{riess22}. If we assume $\fbin=50\%$ in the reference galaxy, then the maximum shift in a distance modulus to any target galaxy is around $\pm 0.004$~mag in $W_H$. If we further assume that the binary percentages in the 37 target galaxies are distributed uniformly between 0 and 100\%, then the 37 shifts in distance moduli will also create a uniform distribution, with a mean of 0~mag and a standard deviation of 0.002~mag. This means that the binarity of Cepheids does not change the value of $H_0$, but does increase the statistical error on $H_0$ by $0.07\,\kmsMpc$ or 0.1\%\footnote{
We estimated the change in $H_0$ from Eq. 2.39 of \citet{breuval21PhD}: $H_0'/H_0 = 10^{-0.2 \Delta \mu}$, where $H_0$ is the most recent value of the Hubble-Lema\^{i}tre constant, $73.04\pm1.04~\kmsMpc$ \citep{riess22}, and $H_0'$ is the value of $H_0$ changed by $\Delta \mu = 0.002$\,mag. We consider the difference $|H_0'-H_0|=0.07\,\kmsMpc$ as the statistical error on $H_0$.
}. The values presented above are much smaller than the uncertainty of the most accurate fit of the PLR zero point derived so far \citep[0.017~mag in $W_H$,][]{cruz-reyes22}, meaning that in $W_H$ \emph{the effect of binarity is insignificant and stays below the detection threshold set by the uncertainty of the PLR zero point fit}.

\section{Discussion}
\label{sec:discussion}

\subsection{Comparison with previous studies}
\begin{figure}
    \centering
    \includegraphics[width=0.95\columnwidth]{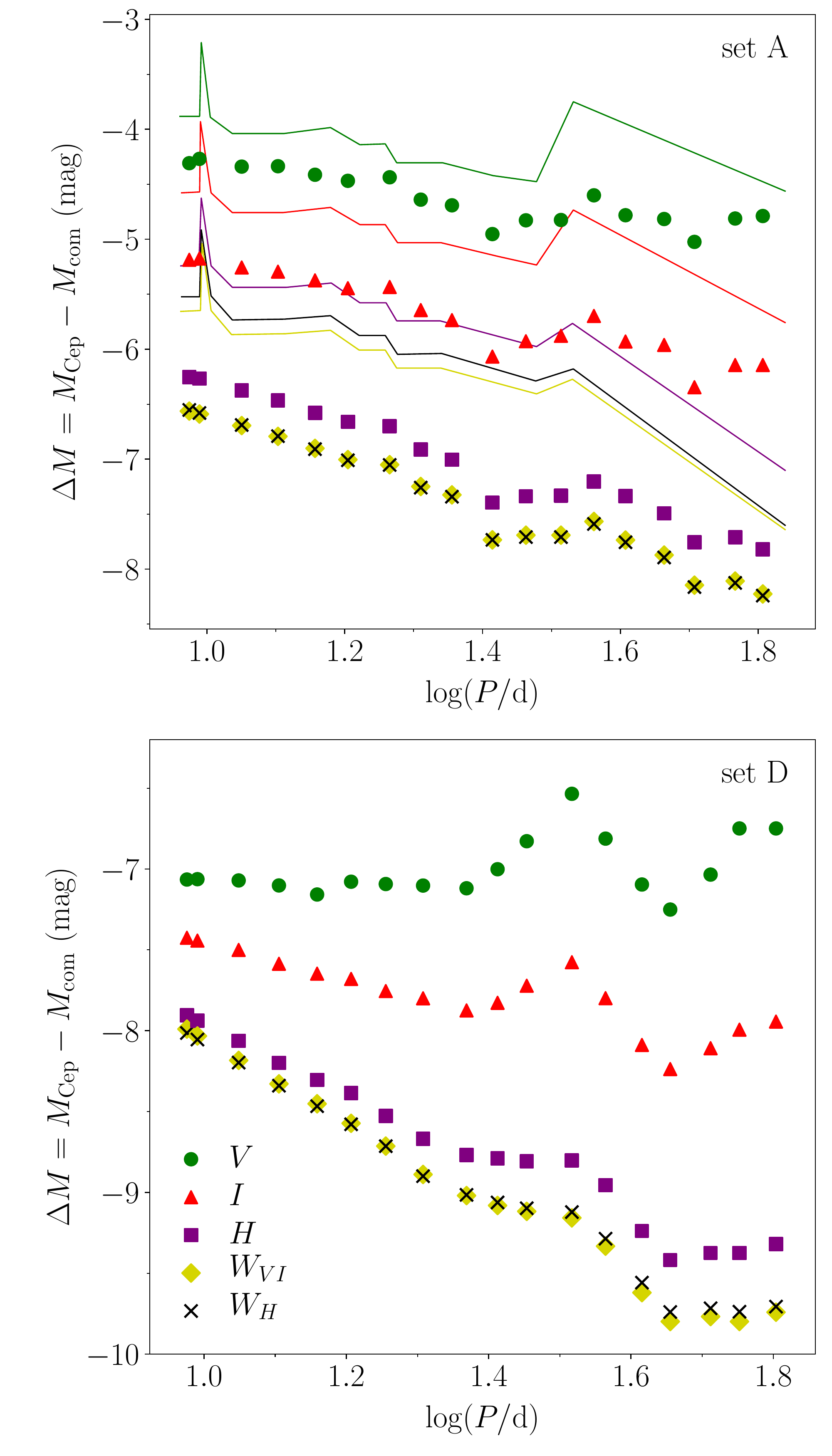}
    \caption{The magnitude difference between Cepheids and their companions in selected photometric passbands. The presented variants have metallicities $Z = 0.02$, Anderson’s prescription of the IS, and the SFH based on the age distribution of the Cepheids, but different sets of the initial parameter distributions: set A (top) and D (bottom). A direct comparison with \citet[][their Figure 8]{Anderson16ApJS} as solid lines of corresponding colors is presented on the top panel.}
    \label{fig:Delat_mag_vs_logP}
\end{figure}

The need to perform population synthesis in order to assess the effect of Cepheid companions on the zero point of the PLR was expressed by \citet{Anderson16ApJS}. They estimated this effect for a Cepheid of $\log P_\mathrm{pul} \sim 1.3$ with a companion which is 6 mag fainter in $W_H$. The brightness difference between a Cepheid and its companion as a function of pulsation period was calculated in $V$, $I$, $H$, $W_{VI}$, and $W_H$, using a number of simplifications, i.a. a fixed mass ratio $q=0.7$; hence the need for a more detailed approach with population synthesis performed here. In Figure \ref{fig:Delat_mag_vs_logP} we compare the brightness difference from \citet{Anderson16ApJS} with our results based on two variants of synthetic populations. The presented variants have metallicities $Z = 0.02$, Anderson’s prescription of the IS, and the SFH based on the age distribution of the Cepheids, but different sets of the initial parameter distributions: set A (top panel) and D (bottom panel). The values in the figure were calculated as a moving median with $\log P$ bins of width 0.15 and with a step of 0.05. The top panel of Figure \ref{fig:Delat_mag_vs_logP} shows a growing discrepancy with increasing wavelength between the results of \citet[][their Figure 8]{Anderson16ApJS} and ours, while in the bottom panel the discrepancy is so large that the comparison lines are not visible anymore. Our synthetic population yielded much smaller values of the brightness difference, i.e., the average companion is much less luminous than the one assumed by \citet{Anderson16ApJS}, which also translates to a much smaller shift of the zero point of the PLR than the value 0.004 mag (for an individual star in $W_H$) they reported. This effect is sensitive to the distribution of mass ratios of companions to Cepheids (as already concluded in Section \ref{sec:differences}).

\citet{anderson18} investigated how Cepheids in wide binaries and open clusters alter distance measurements due to the blending and crowding effects, caused by the limited spatial resolution, especially for galaxies farther away. They investigated wide binary Cepheids\footnote{
A binary is considered wide if its relative semimajor axis is $400 \lesssim a_\mathrm{rel} \lesssim 4000$~au, or equivalently $8.6 \cdot 10^4 \lesssim a_\mathrm{rel} \lesssim 8.6 \cdot 10^5 \,\rsun$ \citep{anderson18}.}
because components of such systems are close enough to be physically bounded, yet far enough to be spatially resolved in the Milky Way when observed with the Hubble Space Telescope (HST). Cepheids with physical companions (either in binaries or open clusters) in more distant galaxies fall below the resolution threshold of the HST, thus their binarity or association with open clusters cannot be resolved. \citet{anderson18} found that in the NIR and Wesenheit indices these blending and crowding effects cause 0.004\% and 0.23\% overestimation of $H_0$ due to Cepheid binarity and cluster associations, respectively. 

A number of differences between our and \citeauthor{anderson18}'s approach to determine the effect of binary Cepheids can be pointed out in order to explain the discrepancy between the two results. They adopted, after \citet{moe17}, the fraction of MW binary Cepheids on wide orbits $f = 0.15$, while we allowed the binarity percentage to vary between 0 and 100\%. This freed us from the assumption that the binary fractions in other galaxies are the same as in the MW, and allowed us to explore more possibilities. Next, \citeauthor{anderson18} assumed that all Cepheid companions are MS stars, while our simulations also yielded red giant, and even Cepheid companions, which is consistent with observations \citep[e.g.][]{pilecki21}. The mass range for Cepheid progenitors on the zero age main sequence, $5-9~\msun$ \citep{anderson18}, was extended by us to $3-12~\msun$, alongside the range of orbital periods from \citeauthor{anderson18}'s $\log P_\mathrm{orb} \in [6.5,~7.5]$ to our $\log P_\mathrm{orb} \in [0.5,~7.5]$ (see Figure A1 in \citetalias{karczmarek22}). Although all the above factors contributed to the quantitative difference between our and \citeauthor{anderson18}'s results, we agree that the smallest effect of binary Cepheids is anticipated in the infrared domain and Wesenheit indices, of order of a few millimagnitudes or smaller.

\subsection{Initial Mass Function}
Our collection of synthetic populations was designed to represent various circumstances under which binary Cepheids are created. We focused on four features that are vital for the characteristics of Cepheids and their companions: the shape of the IS (2 variants), the SFH (2 variants), the initial parameter distributions (4 variants), metallicity (3 variants), and we expounded the results in Section \ref{sec:differences}. However, in all variants of synthetic populations a universal IMF was assumed \citep[][see \citetalias{karczmarek22}, Section 2.1]{kroupa01}, even though they vary with metallicity. This is a common practice, despite recent observational evidence that the value of the IMF slope at the high-mass end\footnote{
The IMF slope changes with mass, following the formula of \citet{kroupa01}: $(m) \propto m^{-\alpha_i}$, where $i$ indicates three mass ranges, and specifically $i=3$ relates to massive stars ($> 1 \msun$), for which $\alpha_3=2.35$.
} decreases from its canonical value ($\alpha_3=2.35$) with decreasing metallicity \citep{marks12,martin-navarro15,jerabkova18}. The most metal-poor galaxy in our synthetic population has $\mbox{[Fe/H]} = -0.7$ dex, for which $\alpha_3 \approx 2.2$, marking only a slight change of about 0.1 \citep[][their Figure 4]{marks12}. We therefore consider the change of the IMF with metallicity negligible for the results of our study. Moreover, the value of $\alpha_3=2.35$ stays constant for metallicities above $\mbox{[Fe/H]} = -0.5$ dex \citep{marks12}, which is relevant for the determination of $H_0$, as the SNeIa host galaxies tend to be more metal rich than the anchor galaxies \citep[][their Table 3]{riess22}, meaning that the IMF can be considered metallicity-independent in the broader context of the extragalactic distance scale.

Another factor that can alter the slope of the IMF is the star formation rate (SFR). This factor is irrelevant for a single, isolated star formation event in a gravitationally collapsing region in a molecular cloud because its timescale is relatively short ($\sim 1$\,Myr), but becomes increasingly important for galaxy-wide IMF (gwIMF), which is an integration over stellar IMFs in the entire galaxy \citep{jerabkova18}. In fact, the SFR is critical to the gwIMF slope at the high-mass end: for low SFR ($10^{-5} \msun \mathrm{yr}^{-1}$) the slope is smaller than its canonical value (massive stars are underrepresented), whereas for high SFR ($10^{5} \msun \mathrm{yr}^{-1}$) the slope is larger than its canonical value (massive stars are overrepresented); the SFR of $1 \msun \mathrm{yr}^{-1}$ yields the canonical value of the slope; see \citet[][their Fig. 2]{jerabkova18}.

The interplay between the metallicity and SFR can render unique gwIMF for each and every galaxy, which is non-trivial to examine. For example, the LMC is expected to have a lower value of the gwIMF slope at the high-mass end, due to its low SFR  \citep[$\sim 0.3 \msun \mathrm{yr}^{-1}$ over a time span relevant to the evolution of Cepheids, i.e. $0-200$\,mln yr;][]{harris09} and low metallicity. However, the empirical data are not conclusive, as they yield values from $-2.3$ \citep[i.e. consistent with the canonical value;][]{weisz15} to $-2.8$ \citep{parker98}. This example clues in about complex interplay between processes that shape the gwIMF and non-homogeneous methods of its examination.

We ran an additional set of simulations with the IMF slope of $\alpha_3 = -2.8$, and noticed that the shifts in the zero points of the PLRs, caused by the extra light from the Cepheids' companions, are slightly smaller (i.e. more negative) when compared to the analogous population created with the IMF slope of $\alpha_3 = -2.35$, but the results from both sets are consistent within errors. We therefore conclude that the IMF slope at the high-mass end, ranging from $-2.8$ to $-2.35$, does not affect our results.

\subsection{Initial mass ratios and orbital periods}
Up to date, the effect of metallicity on the distributions of initial mass ratios and orbital periods of binary systems remains unresolved. Our best approach to account for these unknowns when creating synthetic populations was to include all relevant variants of initial distributions (assembled in sets A, B, C, D) regardless of the possible correlations with metallicity. We expect that our collection of synthetic populations is diverse enough to encompass the results that might have been produced if the potential link between the initial parameters and metallicity had been established.

\subsection{Binary population synthesis codes}
Although we strived to diversify our synthetic populations, our results might still be inaccurate as they all are a product of one binary population synthesis code, \ST. To remedy this, the study we conducted here could be repeated using a different code. However, such undertaking might prove superfluous, because the comparison of four binary population synthesis codes (including \ST) showed that their outputs were in good agreement \citep{toonen14}.

\subsection{Binarity fraction of Cepheids}
\citet{moe17} parametrized the binarity fraction as a function of the orbital period and primary mass, based on extensive data of binary stars in the MW. Such a study for other galaxies with different metallicities, like the Magellanic Clouds, is yet to be done. Our analysis of synthetic binary Cepheids could have been further refined by implementing the binarity fraction of \citeauthor{moe17}, but for the sake of consistency, we would have to extrapolate the same binarity fraction on systems in more metal-poor environments. Instead, we conservatively assumed a uniform distribution of binaries for all metallicities, orbital periods and primary masses, which was further scaled with $\fbin$. This allowed us not only to compare different environments, but also investigate the change in the PLR for different values of $\fbin$.

\subsection{Multiplicity of Cepheids}
The final remark about the limited potential of our synthetic binary populations is related to recent studies on multiplicity. \citet{moe17} reported that massive stars ($\geq 5 \msun$) have on average $1.3-1.6$ companions (depending on the primary's mass, see their Table 13), meaning that an average Cepheid has more than one companion. \citet{dinnbier22} hint at a significant fraction of B-type stars that formed in triples and quadruples. Such systems are most probably hierarchical with a satellite low-mass tertiary on a wide outer orbit \citep{evans20}, and therefore they should contribute even less to the light from the Cepheids than the binary companions. However, the effect of triple and quadruple Cepheids on the extragalactic distance scale is a relevant topic for future exploration.

\section{Conclusions}

Binary Cepheids are an inherent and irremovable part of the PLR calibration. Since they appear brighter than their single counterparts (due to the extra---and not accounted for---light of their companions), the zero point of the PLR is shifted toward brighter magnitudes. This shift scales linearly with the binary Cepheid fraction and decreases with increasing wavelength. We confirm the finding of \citet{anderson18}, that in the NIR domain and Wesenheit indices the effect of binary Cepheids is the smallest.

The shifts of the PLR zero points that occur in a pair of galaxies cause a shift in the distance modulus difference between them. The sign of this shift can be positive or negative, depending on the fractions of binary Cepheids in the two galaxies. If they have similar fractions, the shift in the distance modulus is consistent with zero. On the contrary, a maximum shift in the distance modulus is expected when the fractions of binary Cepheids in the two galaxies are 0 and 100\% or vice versa.

The value of the shift depends on the variant of the synthetic population (a unique combination of metallicity, star formation history, shape and location of the instability strip, as well as the distribution of the initial parameters). In the $W_H$ index, used by \citet{riess22} to derive $H_0$, the distance modulus between a pair of galaxies is shifted due to binary Cepheids by $\pm 0.004$ mag at most. Considering the sample of 37 SNeIa host galaxies from \citeauthor{riess22}'s study, whose distance moduli are altered by any value in range $[-0.004, 0.004]$ mag, we concluded that the effect of binary Cepheids does not shift the value of $H_0$, instead it increases the statistical error on $H_0$ by $0.07\,\kmsMpc$ or 0.1\%.

The effect of binary Cepheids on the extragalactic distance scale has proved minuscule yet not negligible, if one aims to achieve a sub-percent precision of $H_0$. The most important step that can be taken to minimize this effect is to observationally establish the percentage of binary Cepheids and the types of Cepheid companions in anchor and SNeIa host galaxies.

\acknowledgments
We thank the anonymous referee, whose pertinent comments helped to improve this paper. P.K. acknowledges constructive comments and insightful discussions with Richard Anderson and Alex Gallenne. The research leading to these results has received funding from the European Research Council (ERC) under the European Union's Horizon 2020 research and innovation program (grant agreements No. 695099 and No. 951549). We also acknowledge support from the Polish Ministry of Science and Higher Education grant DIR/WK/2018/09, the Polish National Science Centre grants BEETHOVEN 2018/31/G/ST9/03050, MAESTRO 2017/26/A/ST9/00446, and BASAL Centro de Astrofisica y Tecnologias Afines BASAL-CATA grant AFB-170002. K.B. acknowledges support from the Polish National Science Centre grant MAESTRO 2018/30/A/ST9/00050. R.S. acknowledges support from the Polish National Science Centre grant SONATA BIS 2018/30/E/ST9/00598. W.G. acknowledges support from the ANID BASAL project ACE210002. This paper is part of the Araucaria Project, an international collaboration, whose purpose is to provide an improved local calibration of the extragalactic distance scale out to distances of a few megaparsecs (\url{araucaria.camk.edu.pl}).

\section*{Data availability}
The data used to create Figures \ref{fig:DMshift}-\ref{fig:DeltaDM_sets} are available at \url{araucaria.camk.edu.pl/index.php/synthetic-population-of-binary-cepheids/}. The collection of synthetic populations is available upon request.


%

\vspace{5mm}


\software{
StarTrack \citep{belczynski02,belczynski08}, SciPy \citep{2020SciPy-NMeth}, NumPy \citep{2020NumPy-Array}, Pandas \citep{pandas}, Jupyter \citep{ipython}, Colab \citep{bisong19}. All images were proudly made with Matplotlib \citep{matplotlib}.
}




\bibliography{mybib}
\bibliographystyle{aasjournal}



\end{document}